# A Meta-Cognitive Swarm Intelligence Framework for Resilient UAV Navigation in GPS-Denied and Cluttered Environments

Fuqiang Lu [1], Mathias Mankoe [1*], Hualing Bi [1], Abdul-Salam Sibidoo Mubashiru [2]

[1] School of Economics and Management, Yanshan University, Qinhuangdao 066004, China

[2] Kwame Nkrumah University of Science and Technology

* Corresponding author

**Abstract**

Autonomous navigation of UAV swarms in perceptually-degraded environments, where GPS is unavailable and terrain is densely cluttered, presents a critical bottleneck for real-world deployment. Existing optimization-based planners lack the resilience to avoid catastrophic convergence to local optima under such uncertainty. Inspired by principles of computational meta-cognition, this paper introduces a novel swarm intelligence framework that enables a fleet of UAVs to autonomously sense, adapt, and recover from planning failures in real-time. At its core is the Self-Learning Slime Mould Algorithm (SLSMA), which integrates three meta-cognitive layers: a situation-aware search strategy that dynamically selects between exploration and exploitation based on perceived search stagnation; a collective memory mechanism that allows the swarm to learn from and avoid previously failed trajectories; and an adaptive recovery behavior that triggers global re-exploration upon entrapment. We formulate the multi-UAV trajectory problem as a resilient planning challenge, with a cost function that penalizes not only path length and collisions but also navigational uncertainty and proximity to failure states. Extensive simulations in synthetically complex 3D worlds and against the CEC 2017 benchmark suite demonstrate the framework's superior performance. The SLSMA does not merely optimize paths; it generates resilient trajectories, demonstrating a 99.5% mission success rate and significantly outperforming state-of-the-art metaheuristics in recovery speed and solution reliability. This work provides a foundational step towards truly autonomous swarms capable of persistent operation in denied and dynamic environments.

# 1. Introduction

The deployment of unmanned aerial vehicles (UAVs) has catalyzed a paradigm shift across numerous fields, from disaster response and precision agriculture to infrastructure inspection and logistics, underpinned by their unparalleled operational versatility in complex, unstructured environments [1]–[5]. The coordination of multiple drones as a cohesive swarm further amplifies this potential, enabling superior efficacy in large-scale missions through distributed sensing and parallel task execution [6]. The linchpin of such cooperative autonomy, however, is effective and resilient flight trajectory planning [7]. The generation of optimal flight paths is not merely a matter of efficiency; it is critical for ensuring operational safety, mitigating inter-agent collision risks, and guaranteeing mission success in the face of stringent physical and environmental constraints [8], [9].

Path planning methodologies are historically bifurcated into traditional deterministic algorithms and modern meta-heuristics. Foundational algorithms such as A* and Dijkstra provided robust solutions for simplified, discrete search spaces [11]–[13]. However, the reality of multi-UAV trajectory optimization in three-dimensional space constitutes a profoundly complex challenge, characterized by high dimensionality, non-linearity, and multi-modality, with competing objectives such as fuel efficiency, flight time, and dynamic threat avoidance [14], [15]. The computational intractability faced by deterministic methods in these landscapes has precipitated a decisive pivot towards stochastic, population-based meta-heuristic algorithms, which are inherently better suited for navigating such complex, noisy search spaces [16].

Drawing inspiration from natural phenomena, meta-heuristics emulate biological or physical systems to conduct robust global searches, making them a predominant choice for UAV path planning [17]–[19]. This has led to the successful adaptation of a diverse array of algorithms, including enhanced Gray Wolf Optimization [21], hybrid Particle Swarm Optimization [22], ant colony systems [23], and multi-swarm fruit fly optimization [24], among others [25]–[30]. Within this pantheon, the Slime Mould Algorithm (SMA) has emerged as a notable contender, praised for its rapid convergence and structural elegance in solving complex optimization problems [31], [32]. Subsequent enhancements, such as the hybrid SMA (HSMA) for multi-objective planning [33], an improved SMA (ISMA) for collaborative scenarios [34], and a hybrid SMA with Cauchy mutation (AHCSMA) [35], have further cemented its relevance

Notwithstanding these advances, a critical gap persists. The standard SMA and its existing variants remain fundamentally reactive, lacking the cognitive capacity to dynamically assess and adapt their search strategy in response to the problem landscape. This manifests as a fragile balance between exploration and exploitation, a pronounced susceptibility to premature convergence, and an inability to autonomously recover from local optima

entrapment—shortfalls that are particularly debilitating in the high-stakes context of UAV swarm navigation in GPS-denied and cluttered environments.

To bridge this gap, this paper introduces a Meta-Cognitive Swarm Intelligence Framework, instantiated through a novel Self-Learning Slime Mould Algorithm (SLSMA). We posit that true resilience in autonomous navigation requires algorithms that do not merely search, but can learn from the search process itself. The principal novelty of this work lies in the introduction of an algorithm architecture endowed with self-directed learning capabilities, founded on three core innovations:

1) A Situation-Aware Search Strategy: We engineer a novel search mechanism that synergistically integrates the exploration prowess of a ranking-based differential evolution (Rank-DE) with the exploitation focus of the SMA. This allows the algorithm to dynamically and precisely modulate its behavior based on perceived search progression and stagnation.

2) Autonomous Diversity Management: The framework incorporates a dynamic switching operator and an adaptive meta-perturbation technique, enabling it to autonomously maintain population diversity and execute targeted escapes from local optima, thereby accelerating convergence.

3) A Formulation for Resilient Navigation: We rigorously formulate the multi-drone trajectory planning problem within a high-fidelity, complex 3D terrain model, defining it as a constrained optimization task with a holistic cost function that penalizes path length, altitude, maneuver complexity, and—critically—collision risk and navigational uncertainty.

The performance of the proposed SLSMA is rigorously validated against the conventional SMA and other state-of-the-art metaheuristics using the Congress on Evolutionary Computation (CEC) 2017 benchmark suite and demanding multi-drone flight scenarios. The results establish our framework not as a mere improvement, but as a foundational step towards resilient autonomy for UAV swarms

The remainder of this paper is structured as follows: Section 2 details the problem formulation for multi-drone path planning. Section 3 reviews the foundational principles of the traditional SMA and Rank-DE. Section 4 provides a comprehensive mathematical exposition of the proposed SLSMA. Section 5 presents a thorough analysis of the experimental results, and Section 6 concludes the paper and outlines promising future research directions.

## 2. Statement of the Research Problem

### 2.1. Problem Scenario Settings

To mimic a real complex terrain environment, a complex terrain model which includes a base and an obstacle terrains is simulated. Adopted from Nikolos et al [38], the mathematical formulation for the base terrain model is presented in equation (1).

$$Z_1(x, y) = \sin(y + a) + b \cdot \sin(x) + c \cdot \cos\left(d \cdot \sqrt{(y^2 + x^2)}\right) + e \cdot \cos(y) + f \cdot \sin\left(f \cdot \sqrt{(y^2 + x^2)}\right) + g \cdot \cos(y) \quad (1)$$

where $(x, y)$ denotes point coordinates on the horizontal plane. The rough terrain surfaces are simulated using the coefficients $a, b, c, d, e, f$, and g. These values are constant and can be used to model different types of surfaces. Following some common settings in existing literature [39], [40], this paper applies the following settings $a = 3\pi, b = 1/10, c = 9/10, d = 1/2, e = 1/2, f = 1/2, 3/10$.

Whereas the base terrain model focuses on the relief of the terrain, the obstacle terrain model simulates different obstacle forms. Adopted from Zhang et al [41], the mathematical formulation for the base terrain model is shown in equation (2).

$$Z_2(x, y) = \sum_{n=1}^{N_{obs}} H_n \cdot exp\left[-\left(\frac{x - x_{c,n}}{x_{s,n}}\right)^2 - \left(\frac{y - y_{c,n}}{y_{s,n}}\right)^2\right] \quad (2)$$

where $N_{obs}$ is the count of obstacle, $H_n$ is the height of the $n$-th obstacle. The abscissa and ordinate of the centroid of the $n$-th obstacle are respectively represented by $x_{c,n}$ and $y_{c,n}$. The gradients of the $n$-th obstacle along the X-axis and Y-axis are respectively denoted by $x_{s,n}$ and $y_{s,n}$

A combination of different parameter settings in the aforementioned base and obstacle terrain model yields different complex terrain models. The complex terrain model could be expressed as in equation (3)

$$Z(x, y) = max[Z_1(x, y), Z_2(x, y)] \quad (3)$$

where $Z_1(x, y)$ and $Z_2(x, y)$ denote the heights of the base terrain model and the obstacle terrain model at $(x, y)$ respectively

### 2.2. Representation of Drone Flight Path

The path planning problem in this paper involves the mapping of the flight trajectory, which is composed of waypoint series, of multiple drones using a set of predefined routes. Suppose $M$ is the number of drones, and $N$ is the number of waypoints. Suppose that each drone, $m \in \{1, 2, .., M - 1, M\}$, at the same speed, flies from a starting point, $P_s^m = (x_s^m, y_s^m, z_s^m)$, and lands at a terminal point, $P_t^m = (x_t^m, y_t^m, z_t^m)$. Then, equation (4) is a formulation of the flight trajectory of the drone as a discrete sequence of coordinate points.

$$WP^m = \{P_1^m, P_2^m, P_3^m, \ldots, P_{N-1}^m, P_N^m\}, m = 1, \ldots, M \quad (4)$$

where $P_i^m$ represents waypoint $i$ of drone $m$, $P_1^m = P_s^m$ and $P_N^m = P_t^m$. Equation (5) is a mathematical formulation for each waypoint, $P_i^m$.

$$P_i^m = \{z_i^m, y_j^m, z_i^m\}, i = 1, 2, \ldots, N - 1, N \tag{5}$$

where $x_i^m$, $y_i^m$ and $z_i^m$ represent the abscissa, ordinate, and vertical coordinates in 3D space.

The multi-drone path planning seeks to find a set of suitable flight paths, waypoints, $WP = (WP_1, WP_2, \ldots, WP_{m-1}, WP_M)$ to minimize the total cost of each flight trajectory. Nonetheless, the connecting waypoints do not meet the requirements of drone navigation since the resultant trajectory is non-differentiable though continuous. A smoothing strategy can be used to ensure that the flight trajectories are continuously differentiable and kinematically feasible. Some of the commonly used methods include the B-spline curve, Dubins curve, and the Savitzky-Golay filter [42], [43], [44]. The B-spline curve method is widely used for smooth drone path planning due to its affine invariance, invariant geometry and preserved convexity [45]. Additionally, it provides a significant reduction in the complexity of the smoothing process by keeping the degree of the polynomial unchanged even following addition of more control points. Equation (6) is a mathematical definition for the B-spline curve.

$$P(t) = \sum_{i=0}^{N_{ctrl}} d_i \cdot N_{i,k}(t) \tag{6}$$

where the control points are represented by $d_i, i = 1, 2, \ldots, N_{ctrl} - 1, N_{ctrl}$. The order of the B-spline curves' segments is $k$, while $N_{i,k}(t)$ is the $k$-order normalized blending function described by $(t_0 \leq t_1 \leq \ldots t_{n+k})$. Equation (7) defines the $N_{i,k}(t)$ function.

$$\begin{cases} N_{l,1}(t) = \begin{cases} 1, if\ t_l \leq t \leq t_{l+1} \\ 0, \quad otherwise \end{cases} \\ N_{l,k}(t) = \dfrac{t - t_l}{t_{l+k-1} - t_l} \times N_{l,k-1}(t) + \dfrac{t_{l+k} - t}{t_{l+k} - t_{l+1}} \times N_{l+1,k-1}(t) \end{cases} \tag{7}$$

In this paper, the cubic B-spline curves are utilized to make the flight trajectories of the drones smooth. According to Hasan et al [46], the cubic B-spline curves are highly precise in smoothing complex geometries.

### 2.3. Cost Function

The path planning cost function is a mathematical model that accounts for the various factors that affect the flight path. It also accounts for the constraints of the trajectories. This section presents the five key factors that are incorporated into the cost function.

### 2.3.1. Cost of Drone Flight Distance

The flight time of drones is an important factor. A less flight time allows them to reach their intended location more quickly and consume less energy. To ensure a less flight time is equivalent to reducing the distance flown if the drone flies at a steady speed. Equation (8) is a set of connected waypoints, representing drone flight distance.

$$D^m = \{\|\overrightarrow{P_1^m P_2^m}\|, \ldots, \|\overrightarrow{P_{N-1}^m P_N^m}\|\} \tag{8}$$

where $D^m$ represents the $m$-th drone's set of flight segments. $\|\overrightarrow{P_i P_j}\|$ represents the Euclidean distances from waypoint $P_i$ to waypoint $P_j$. Equation (9) is the mathematical formula for calculating the cost of flight distance for drone $m$.

$$C_{dist}^m = \frac{\sum_{n=1}^{N-1} \|\overrightarrow{P_n P_{n+1}}\|}{\|\overrightarrow{P_1 P_N}\|} \tag{9}$$

**2.3.2. Cost of Drone Flight Height**

It is also important to optimize the altitude variations to minimize energy consumption of drones during flight. The drones must operate within some defined height ranges. Equations (10) and (11) are the mathematical formula for calculating the cost of flight height for drone $m$.

$$C_{height}^m = \sum_{i=2}^{N-1} Q_{height,i}^m \tag{10}$$

$$Q_{height,i}^m = \begin{cases} q_{height}, & \text{if } z_i^m < z_{lb} \text{ or } z_i^m > z_{ub} \\ 0, & \text{otherwise} \end{cases} \tag{11}$$

For drone $m$ at waypoint $i$ with a height of $z_i^m$, the costs of flight height is denoted by $Q_{height,i}^m$. For all the drones, the lower and upper flight height boundaries are represented by $z_{lb}$ and $z_{ub}$ respectively. The flight height has a cost coefficient of $q_{height}$.

**2.3.3. Cost of Drone Flight Turning**

Drones can only turn within a specified angle range in both vertical and horizontal directions. Any maneuver that exceeds these limits is considered infeasible, which is why restricting the pitch and yaw angles during flight is necessary. Equations (12), (13), and (14) are the mathematical formulae for computing the cost of flight turning for drone $m$.

$$C_{turn}^m = \sum_{i=2}^{N-1} Q_{yaw,i}^m + \sum_{1=2}^{N-1} Q_{pitch,i}^m \tag{12}$$

$$Q_{yaw,i}^m = \begin{cases} q_{yaw}, & \text{otherwise} \\ 0, & \text{if } 0 \leq \alpha_i^m \leq \alpha_{max} \end{cases} \tag{13}$$

$$Q_{pitch,i}^m = \begin{cases} q_{pitch}, & \text{otherwise} \\ 0, & \text{if } 0 \leq \beta_i^m \leq \beta_{max} \end{cases} \tag{14}$$

For drone $m$, the cost of yaw angle $i$ is represented by $Q_{yaw,i}^m$. Similarly, the cost of pitch angle $i$ is represented by $Q_{pitch,i}^m$. $\alpha_i^m$ represents yaw angle $i$, $\beta_i^m$ represents pitch angle $i$. The costs of flight turning have coefficients of $q_{yaw}$ and $q_{pitch}$. Adopted from Xiong et al [47] and Zhang et al [48], equations (15) and (16) are the mathematical formulae for computing the yaw and pitch angles respectively.

$$\alpha_i^m = arccos\left(\frac{(x_i^m - x_{i-1}^m)(x_{i+1}^m - x_i^m) + (y_i^m - y_{i-1}^m)(y_{i+1}^m - y_i^m)}{\sqrt{(x_i^m - x_{i-1}^m)^2 + (y_i^m - y_{i-1}^m)^2} \cdot \sqrt{(x_{i+1}^m - x_i^m)^2 + (y_{i+1}^m - y_i^m)^2}}\right) \quad (15)$$

$$\beta_l^m = arctan\left(\frac{|Z_l^m - Z_{l-1}^m|}{\sqrt{(x_l^m - x_{l-1}^m)^2 + (y_l^m - y_{l-1}^m)^2}}\right) \quad (16)$$

**2.3.4. Cost of Collision Between Drones and other Obstacles**

It is imperative to ensure the safety of drones during a mission. Collisions with obstacles such as rocks, skyscrapers, and mountains could be costly. The cost of drone $m$ colliding with some obstacles is computed using equations (17) and (18).

$$C_{collis\_obs}^m = \sum_{i=1}^{N-1} Q_{collis\_obs, i}^m \quad (17)$$

$$Q_{collis\_obs, i}^m = \begin{cases} q_{collis\_obs}, & \text{if } (x_i^m, y_i^m, z_i^m) \text{ is in obstacles} \\ 0, & \text{otherwise} \end{cases} \quad (18)$$

For drone $m$, at waypoint $i$, the cost of colliding with an obstacle is denoted by $Q_{collis\_obs, i}^m$ with a coefficient of $q_{collis\_obs}$.

**2.3.5 Cost of Collision Between Drones**

When multiple drones fly at the same time, it is important to ensure no drone-drone collisions too. These mishaps can occur when the flight paths overlap or the distances between them is poor, which can lead to significant system malfunctions. Equations (19) and (20) are the formulae for computing the cost of collision between drone $m$ and other drones.

$$C_{collis\_drone}^m = \sum_{i=2}^{N-1} \sum_{n=1, n \neq m}^{M} Q_{collis\_drone, i}^{m, n} \quad (19)$$

$$Q_{collis\_drone, i}^{m, n} = \begin{cases} 0, & \text{if } \left\|\overrightarrow{P_i^m P_i^n}\right\| < d_{min} \\ q_{collis\_drone}, & \text{otherwise} \end{cases} \quad (20)$$

For any two drones, $m$ and $n$, at waypoint $i$, the cost of collision is denoted by $Q_{collis\_drone, i}^{m, n}$. The Euclidean distance between them is denoted by $\left\|\overrightarrow{P_i^m P_i^n}\right\|$ with $d_{min}$ as the minimum safe distance and $q_{collis\_drone}$ as the cost coefficient.

**2.3.6. Total Cost**

The five drone flight constraints are written in as cost functions and combined to form a single mathematical function of costs to compute the total cost. Equation (21) is the mathematical model for the total cost for all the drone flights.

$$C_{total} = \sum_{m=1}^{M} (\omega_1 \cdot C_{distance}^m + \omega_2 \cdot C_{height}^m + \omega_3 \cdot C_{turn}^m + \omega_4 \cdot C_{collis\_obs}^m + \omega_5 \cdot C_{collis\_drone}^m) \quad (21)$$

where $\omega_i, i = 1, 2, \ldots, 5$, are the weight coefficients for the respective cost functions.

## 3. Background of Methods

### 3.1. Traditional SMA

The SMA is inspired by the patterns of slime molds that are found in nature. It takes advantage of adaptive weights to model both the negative and positive feedback that occurs during the foraging process. It also abstracts the changes that happen in the foraging environment [49].

#### 3.1.1. Navigation Toward Food

The individual slime moulds in slime mold colonies rely on the level of odour in the air to navigate toward food. Assume that the slime mould has a population size of $NP$, equation (22) is the mathematical formula for updating position of a slime mould.

$$X(t+1) = \begin{cases} X_{best}(t) + v_b \cdot (W \cdot X_A(t) - X_B(t)), & if\ rand_1 < p \\ v_c \cdot X(t), & if\ rand_1 \geq p \end{cases} \quad (22)$$

where the current iteration's index is denoted by $t$, the current individual iteration position is denoted by $X(t)$ and the next is represented by $X(t+1)$, the optimal position is represented by $X_{best}(t)$, $X_A(t)$ and $X_B(t)$ denote two individual slime moulds randomly sampled from the population. $rand_1$ is a number sampled randomly from the Uniform distribution [0, 1]. $p$ denotes a conrol parameter for slime mould position updating patterns. Equation (23) is a mathematical formula for $p$.

$$p = tanh|Fit(X_i) - Fit_{best}|, i = 1, 2, \ldots, NP - 1, NP \quad (23)$$

where the fitness of slime mould $i$ is denoted by $Fit(X_i)$, the overall best fitness is denoted by $Fit_{best}$, the slime moulds' selection behavior when approaching food are simulated by parameters $v_b$ and $v_c$, with the former denoting the extent of slime mould exploration and bounded by $[-a, a]$, and the latter denoting individual slime mould's exploitation of historical information, which declines from 1 to 0. The mathematical formula for $a$ is expressed in equation (24).

$$a = arctanh\left(-\left(\frac{t}{t_{max}}\right) + 1\right) \quad (24)$$

where the maximum number of iterations is denoted by $t_{max}$.

Equation (25) is the mathematical formula for computing the weight of a slime mould, $W$.

$$W(FitIndex(i)) = \begin{cases} 1 + rand_1 \cdot log\left(\frac{BF - Fit(X_i)}{BF - WF} + 1\right), & condition \\ 1 - rand_1 \cdot log\left(\frac{BF - Fit(X_i)}{BF - WF} + 1\right), & others \end{cases} \quad (25)$$

$$FitIndex = sort(Fit) \quad (26)$$

where the sequence determined after assessing the fitness level of the slime molds is represented by $FitIndex$. The sorting process in minimization is conducted in an ascending order. Whether or not $Fit(X_i)$ ranks first part of the population is denoted by $condition$.

Given the current iteration, whereas $BF$ denotes the best fitness, the worst fitness is represented by $WF$.

### 3.1.2. Food Wrapping

The tissue structures of slime mold veins are contracted to wrap food. The concentration of food particles in contact with these structures influences the waves produced by the mold's biological oscillator. A higher concentration leads to more powerful waves, which can increase the speed of the movement of cytoplasm and lead to thicker veins. Slime molds tend to focus on finding an area with high food concentrations to exploit. When the food concentration falls, they search for other sources of food. Equation (27) is the position updating mathematical formula for a slime mould.

$$X(t+1) = \begin{cases} rand_2 \cdot (UB - LB) + LB, & rand_2 < z \\ X_{best}(t) + v_b \cdot (W \cdot X_A(t) - X_B(t)), & rand_1 < p \\ v_c \cdot X(t), & rand_1 \geq p \end{cases} \quad (27)$$

where the lower boundary of the search space is denoted by $LB$, the upper boundary of the search space is denoted by $UB$, $rand_2$ denotes a number sampled randomly from the Uniform distribution $[0,1]$, $z$ denotes a percentage of slime moulds engaged in random exploration, empirically fixed at 0.03.

### 3.1.3. Oscillation

Slime moulds use their bio-oscillators to adjust the flow rate of fluids in their veins based on the concentration of food. This helps them find the ideal food locations by altering the veins' width. The parameters $W$, $v_b$ and $v_c$ work together to control the oscillation mechanism in SMA, which aim to balance the exploitation and exploration.

## 3.2. Ranking-based Differential Evolution (Rank-DE)

As an alternative to the DE, the rank-DE is widely applied to solve global optimization problems [50]. It is based on the concept that the more desirable traits of parent organisms are more likely to result in their offspring [37]. The following subsections present a detailed description of the rank-DE.

### 3.2.1. Assignment of Rankings

The rank-DE system takes into account the entropy characteristic of high-quality individuals and ranks them according to their fitness values. For instance, the best individuals receive a high ranking while the worst receive a low one. Equation (28) is a mathematical formula for assigning the ranking of an individual.

$$R_j = NP - j, j = 1, 2, \ldots, NP - 1, NP \quad (28)$$

where $j$ denotes the ranking index following individuals sorting.

### 3.2.2. Probability of Selection

The selection probability is calculated by considering the individual's rank. Equation (29) is a mathematical formula for computing the selection probability.

$$P_j = \frac{R_j}{NP}, j = 1, 2, \ldots, NP - 1, NP \qquad (29)$$

where the probability of selecting an individual ranked in position j is represented by $P_j$.

### 3.2.3. Selection of Vector

The selection probability of a mutation in the rank-DE determines its operator. This paper uses the DE-Rand1 strategy. The definition for the "DE-Rand1" mutation strategy is presented in Equation (30).

$$V_i = X_{r1} + F \cdot (X_{r2} - X_{r3}) \qquad (30)$$

where $r_i$, $i = 1, 2,$ and 3 are mutually exclusive integers sampled randomly from $[1, NP]$. $F$ is a scaling factor sampled from $[0, 2]$. The current position of individual $i$ and the post-mutation position are denoted by $X_i$ and $V_i$ respectively.

The rank-DE algorithm applied in this paper chooses the some vectors based on their probability. The first and base is $X_{r1}$, while the second and terminal is $X_{r2}$. To enhance the algorithm's randomization, $X_{r3}$ is randomly sampled from the population. Algorithm I shows the pseudo-code for the rank-DE vector selection, which uses the mutation strategy of "DE-Rand1".

| Algorithm I: Rank-DE Vector Selection with Differential Evolution and Rand1 (DE-Rand1) |
|---|
| **Inputs**: Population Size($NP$), Selection Probability ($P$), Target Individual Index ($i$) <br> **Outputs**: Selected Individual Indices, $r_1, r_2, and\ r_3$ <br> 1.   $r_1 \leftarrow$ select randomly from $[1, NP]$; <br> 2.   **while** $rand(0, 1) > P_{r1}$ **or** $r_1 == i$ **do** <br> 3.      $r_1 \leftarrow$ select randomly from $[1, NP]$; <br> 4.   **end while** <br> 5.   $r_2 \leftarrow$ select randomly from $[1, NP]$; <br> 6.   **while** $rand(0, 1) > P_{r2}$ **or** $r_1 == i$ **or** $r_2 == i$ **do** <br> 7.      $r_2 \leftarrow$ select randomly from $[1, NP]$; <br> 8.   **end while** <br> 9.   $r_3 \leftarrow$ select randomly from $[1, NP]$; <br> 10. **while** $r_3 == r_2$ **or** $r_3 == r_1$ **or** $r_3 == r_i$ **do** <br> 11.     $r_3 \leftarrow$ select randomly from $[1, NP]$; <br> 12. **end while** |

Thereafter, $X_i$ and $V_i$ are combined using a crossover operator as shown in equations (31) and (32).

$$U_i = [U_1, U_2, \ldots, U_{Dim-1}, U_{Dim}]^T \qquad (31)$$

$$U_{i, Dim} = \begin{cases} V_{i, Dim}, & if\ rand < C_r\ or\ D_{rand} = Dim \\ X_{i, Dim}, & otherwise \end{cases} \qquad (32)$$

where the dimension of the problem is denoted by $Dim$, $D_{rand}$ represents an integer randomly sampled from $[1, Dim]$, $rand$ denotes a number sampled randomly from $[0, 1]$, and the crossover parameter is denoted by $Cr$.

Then, a greedy selection criterion is applied to keep relatively better individuals. Suppose the new individual ($U_i$) has a better fitness than that of the current individual ($X_i$), then the former is kept. Else, the latter is maintained for the following generation.

**4. The Proposed Algorithm, SLSMA**

The SLSMA is developed and proposed to tackle the limitations of the SMA, particularly for planning multi-drone flight paths. The following three points constitute the new and key improvements incorporated in SMA to better its performance.

1. A new mechanism for searching the search space. The proposed SLSMA combines the Rank-DE's exploration capabilities and a crossover and mutation procedure to address the imbalance in the exploitation and exploration process.

2. A dynamic switch operator. The proposed SLSMA integrates an adaptive probability that changes throughout the iteration process, which improves the optimal solution's quality and prevent it from prematurely converging.

3. A dynamic perturbation strategy. The proposed SLSMA conducts an update of the positions of individuals who are performing poorly when the population encounters a period of convergence stagnation.

Figure 1 depicts the conceptual framework of the proposed SLSMA for solving optimization problems involving multi-drone path planning.

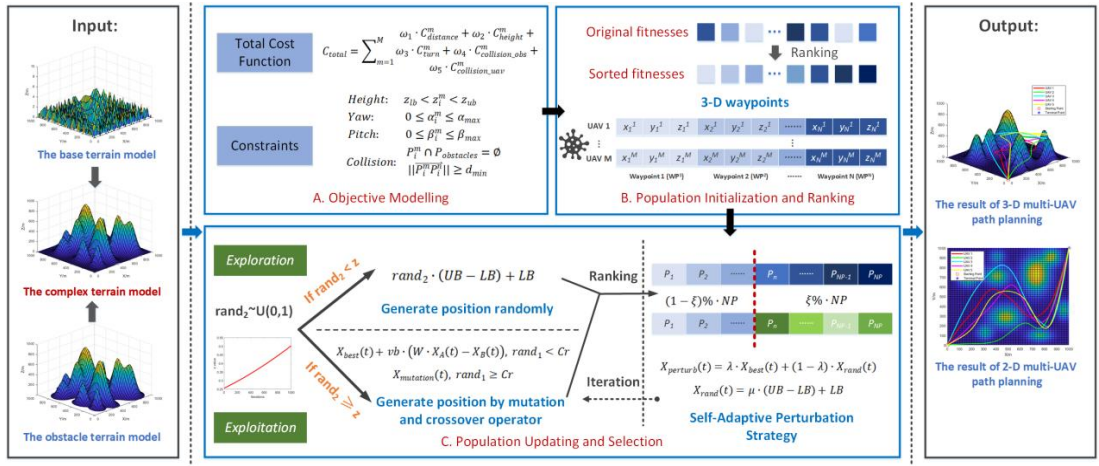

*Figure 1*: A conceptual framework of the SLSMA.

**4.1. A New Search Scheme**

The trade-off between exploration and exploitation is a significant challenge for meta-heuristics during the search process [51]. During the exploration phase, the individuals are more likely to look for solutions that are globally ideal. Nevertheless, over exploration may lead to algorithm convergence issues. During the exploitation phase, on one hand, the individuals tend to search in small steps near the ideal area, which leads to faster population convergence. On the other hand, excessive exploitation can result in premature convergence since the algorithm gets trapped in local optima. In equation (27), for $rand_2 < z$, the algorithm creates positions randomly to simulate the exploration behavior. For $rand_2 \geq z$, the algorithm adjusts the value of slime molds' exploitation range based on $v_b$ and $v_c$. This approach limits the exploration potential of slime molds. In many experiments, it has been observed that the value of optima is typically set to 0.03, which makes it hard for the algorithm to escape local optima. This is especially true when the population gets homogenized.

Therefore, a new mechanism that replaces the exploitation phase in the SMA algorithm is proposed by utilizing the rank-DE mutation operation. The algorithm's position update mode is then changed to accommodate Cr, which is a crossover operator. This combination of the two functions increases the likelihood of finding solutions in the global search.. Equation (33) is the mathematical formula for updating the positions of SLSMA.

$$X_{new}(t+1) = \begin{cases} rand_2 \cdot (UB - LB) + LB, & rand_2 < z \\ X_{best}(t) + v_b \cdot (W \cdot X_A(t) - X_B(t)), & rand_1 < C_r \\ X_{muta}(t), & rand_1 \geq C_r \end{cases} \quad (33)$$

where the new resultant updated position is denoted by $X_{new}(t)$, and the post mutation individual position is denoted by $X_{muta}(t)$, as expressed by equation (34).

$$X_{muta}(t) = X_{r1}(t) + 2 \cdot rand_3 \cdot cb \cdot (X_{r2}(t) - X_{r3}(t)) \quad (34)$$

The scaling factor $F$ is substituted with a number sampled randomly from $[0, 2a]$ in order to prevent new parameters from being introduced in the algorithm. In algorithm I, $r_1, r_2$, and $r_3$ are chosen.

**4.2. A Dynamic Switch Operator**

The complexity and non-linear nature of the optimization processes for real-world problems significantly hinder the performance of fixed-search algorithms. In SMA, the parameter $z$ determines the balance between local exploitation and global exploration. As the population grows and the differences decrease, the value of $z$ challenges better solution exploration capacity. Hence, the SLSMA algorithm converts the parameter $z$ from a fixed to a non-linear one. This allows the algorithm to perform robust global searches. Equation (35) is the mathematical formula for calculating the dynamic switch operator.

$$z = log_{10}\left(1 + 0.8 \times exp\left(\frac{t+1}{t_{max}}\right)\right) \quad (35)$$

The Figure 2 visualizes the dynamic switch operator against iterations. In the early iterations, $z$ is relatively small, which encourages individual slime moulds to explore promising regions. The value of $z$ then rises as the number of iterations goes up, causing the population to move beyond the current to search for other promising alternatives.

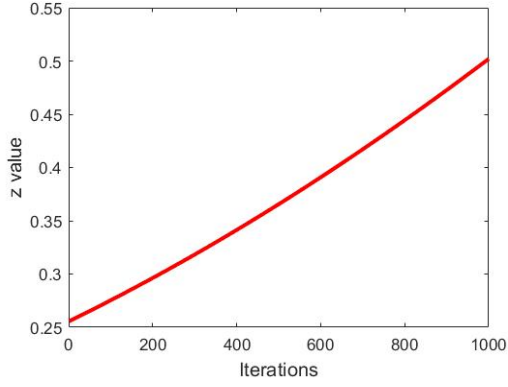

*Figure 2*: Dynamic switch operator.

**4.3. A Dynamic Perturbation Technique**

The position of the population tends to stabilize when it stagnates for many iterations. This leads to the development of algorithms that are unable to find solutions that are better than their current state. By modifying the population structure, algorithms can search for new solutions.

The dynamic perturbation mechanism searches the local optima space to bypass any potential local optima through introduction of minor variations following algorithm stagnation. The SLSMA algorithm's ability to determine if a perturbation strategy is needed is a function of the stagnation counter, which is incremented if the difference of the values of the optimal fitness of the current and previous iterations is less than $\varepsilon = 1.0E - 03$, the perturbation operator. Otherwise, the stagnation counter is set to zero. Where the predetermined index of perturbation, $N_{stag} = 15$, is reached, the algorithm is deemed to have stagnated, which calls for the implementation of the perturbation mechanism.

The dynamic perturbation strategy seeks to improve the population by substituting with new individuals the individuals with poor performance. Given a sign of stagnation, the individuals with worst performance, ξ, are removed, following which new individuals, ξ, are added to the population. The mathematical formula for the dynamic perturbation strategy is shown in equation (36).

$$X_{pert}(t) = \lambda \cdot X_{best}(t) + (1 - \lambda) \cdot X_{rand}(t) \quad (36)$$

$$X_{rand}(t) = \mu(UB - LB) + LB \quad (37)$$

where both $\lambda$ and $\mu$ denote numbers randomly sampled from the Uniform distribution, [0, 1]. Given all iterations, the optimal position is denoted by $X_{best}(t)$. Algorithm II represents the pseudo-code of the SLSMA.

| Algorithm II: Pseudo-Code of Proposed Algorithm, SLSMA |
|---|

**Inputs**: Population Size ($NP$), Dimension ($Dim$), Lower Boundary ($LB$), Upper Boundary ($UB$), Maximum Number of Iterations ($t_{max}$), Crossover Rate ($Cr$), and Perturbation Rate ($\xi$).
**Outputs**: Optimal Position ($X_{optimal}$), Optimal Fitness ($Fit_{optimal}$).

1. Initialize the positions of population $X$ and parameters;
2. $P \leftarrow$ compute using equation (29);
3. $N_{stag\_count} = 0$; $t = 1$;
4. **while** $t \leq t_{max}$ **do**
5.    $z \leftarrow$ compute using equation (35);
6.    $a \leftarrow$ compute using equation (24);
7.    **for** $i = 1: NP$ **do**
8.      $Fit(X_i) \leftarrow$ compute the fitness value of each individual $X_i$;
9.    **end for**
10.   $W \leftarrow$ compute using equation (25);
11.   $X_{best}, Fit_{best} \leftarrow$ evaluate the global optimal position and global optimal fitness;
12.   **for** $i = 1: NP$ **do**
13.     **if** $rand_2 < z$ **then**
14.       $X_{i, new} \leftarrow$ update the new position using equation (33)(1);
15.     **else**
16.       **for** $d = 1: Dim$ **do**
17.         **if** $rand_1 < Cr$ or $D_{rand} = Dim$ **then**
18.           $X_{i, d, new} \leftarrow$ update the new position using equation (33)(2);
19.         **else**
20.           $r_1, r_2$, and $r_3 \leftarrow$ select using Algorithm I;
21.           $X_{i, d, new} \leftarrow$ update the new position using equation (33)(3);
22.         **end if**
23.       **end for**
24.     **end if**
25.     **if** $Fit_{i, new} < Fit(X_i)$ **then**
26.       **replace** $X_i$ with $X_{i, new}$;
27.     **end if**
28.   **end for**
29.   $X_{best}, Fit_{best} \leftarrow$ update the global optimal position and global optimal fitness;
30.   **if** $|Fit_{best, t} - Fit_{best, t-1}| < \varepsilon$ **then**
31.     $N_{stag\_count} = N_{stag\_count} + 1$;
32.   **else**
33.     $N_{stag\_count} = 0$;
34.   **end if**
35.   **if** $N_{stag\_count} \geq N_{stag}$ **then**
36.     $X_{pert} \leftarrow$ implement self-adaptive perturbation strategy using equation (36);
37.   **end if**
38. **end while**

## 5. Results of Simulations and Analyses

This section compares the proposed SLSMA and some seven well-known meta heuristic algorithms on the Congress on Evolutionary Computation 2017 benchmark test suite. In addition, experiments were performed on the path planning problems with three well-established standard SMA variants and the SLSMA. All the simulations, experiments, and analyses conducted in this paper were performed on a Windows 10 operating system with 8GB of RAM, Intel(R) Core (TM) i5-8250U CPU, and a main frequency of 1.80 GHz. The implementation and visualization of the algorithms were done in MATLAB R2024a.

**5.1. Experiments on the Congress of Evolutionary Computation 2017 Test Suite**

The Congress on Evolutionary Computation 2017 benchmark test suite, which is composed of four categories, is highly recognized and widely used to evaluate single objective optimization problems involving a real parameter [52]. In all, the four categories, uni-modal functions $(F_1, F_3)$, simple multi-modal functions $(F_4 - F_{10})$, hybrid functions $(F_{11} - F_{20})$, and composition functions $(F_{21} - F_{30})$ consist of 29 functions defined over the interval [-100, 100]. Tables 1-4 show the details of the 29 functions.

**Table 1: Congress on Evolutionary Computation 2017 Uni-modal Functions**

| Number | Functions | Optimal Fitness Value |
|---|---|---|
| 1 | $F_1(x)$: Shifted and rotated Bent Cigar function | 100 |
| 2 | $F_3(x)$: Shifted and rotated Zakharov function | 200 |

**Table 2: Congress on Evolutionary Computation 2017 Simple Multi-modal Functions**

| Number | Functions | Optimal Fitness Value |
|---|---|---|
| 1 | $F_4(x)$: Shifted and rotated Rosenbrock's function | 300 |
| 2 | $F_5(x)$: Shifted and rotated Rastrigin's function | 400 |
| 3 | $F_6(x)$: Shifted and rotated expanded Scaffer's $F_7$ function | 500 |
| 4 | $F_7(x)$: Shifted and rotated Lunacek Bi-Rastrigin function | 600 |
| 5 | $F_8(x)$: Shifted and rotated non-continuous Rastrigin's function | 700 |
| 6 | $F_9(x)$: Shifted and rotated Levy function | 800 |
| 7 | $F_{10}(x)$: Shifted and rotated Schwefel's function | 900 |

**Table 3: Congress on Evolutionary Computation 2017 Hybrid Functions**

| Number | Functions | Optimal Fitness Value |
|---|---|---|
| 1 | $F_{11}(x)$: Hybrid function 1 (N=3) | 1000 |
| 2 | $F_{12}(x)$: Hybrid function 2 (N=3) | 1100 |
| 3 | $F_{13}(x)$: Hybrid function 3 (N=3) | 1200 |
| 4 | $F_{14}(x)$: Hybrid function 4 (N=4) | 1300 |
| 5 | $F_{15}(x)$: Hybrid function 5 (N=4) | 1400 |
| 6 | $F_{16}(x)$: Hybrid function 6 (N=4) | 1500 |
| 7 | $F_{17}(x)$: Hybrid function 7 (N=5) | 1600 |
| 8 | $F_{18}(x)$: Hybrid function 8 (N=5) | 1700 |
| 9 | $F_{19}(x)$: Hybrid function 9 (N=5) | 1800 |
| 10 | $F_{20}(x)$: Hybrid function 10 (N=6) | 1900 |

**Table 4: Congress on Evolutionary Computation 2017 Composite Functions**

| Number | Functions | Optimal Fitness Value |
|---|---|---|
| 1 | $F_{21}(x)$: Composition function 1 (N=3) | 2000 |
| 2 | $F_{22}(x)$: Composition function 2 (N=3) | 2100 |
| 3 | $F_{23}(x)$: Composition function 3 (N=4) | 2200 |
| 4 | $F_{24}(x)$: Composition function 4 (N=4) | 2300 |

| 5  | $F_{25}(x)$: Composition function 5 (N=5)  | 2400 |
| 6  | $F_{26}(x)$ = Composition function 6 (N=5) | 2500 |
| 7  | $F_{27}(x)$: Composition function 7 (N=6)  | 2600 |
| 8  | $F_{28}(x)$ = Composition function 8 (N=6) | 2700 |
| 9  | $F_{29}(x)$ = Composition function 9 (N=3) | 2800 |
| 10 | $F_{30}(x)$ = Composition function 10 (N=3)| 2900 |

**5.1.1. Analysis of Algorithmic Effectiveness**

The proposed algorithm's effectiveness was evaluated by comparing it with DE [53], GWO [18], PSO [54], SCA [57], SMA [49], SSA [56], and WOA [55] using the Congress on Evolutionary Computation 2017 benchmark test suite. The algorithms were assessed under same and fair conditions. The population size and the problem dimension of the algorithms were set to 30. The algorithms were run independently to minimize the effects of randomness on the experiments, with 1000 maximum number of iterations.

The three evaluation metrics used in this analysis were Average Fitness $(AVG)$, Best Fitness $(BEST)$, and Standard Deviation $(STD)$. They indicate the optimal algorithm for achieving the best results while ensuring the consistency of the approach in several tests. The stability of the algorithm is also taken into account to see if it performs well.

The performance of algorithms was evaluated using the Wilcoxon rank-sum test and the Friedman test. The latter examines the statistical significance of the average ranking differences [58]. The algorithm which ranked first was considered the best. The former test ensures that the best algorithmic performance is not random, but significant statistically [59]. Table 5 details the parameter settings for the algorithms under study. The SLSMA parameters were set through multiple experiments. The other algorithms' parameter settings follow the recommendations found in their original papers.

**Table 5: Algorithm Parameter Settings**

| Number | Algorithm | Parameter settings |
|---|---|---|
| 1 | DE    | $F = 0.5$, $CR = 0.5$ |
| 2 | GWO   | $a$ is decreased linearly from 2 to 0 |
| 3 | PSO   | $c_1 = 2, c_2 = 2, \omega = 0.4$ |
| 4 | SCA   | $a = 2$ |
| 5 | SLSMA | $Cr = 0.5, \zeta = 0.1$ |
| 6 | SMA   | $z = 0.03$ |
| 7 | SSA   | $c_2 \in [0,1], c_3 \in [0,1]$ |
| 8 | WOA   | $a$ is decreased linearly from 2 to 0, $b = 1$ |

The results of the experiments are presented in Tables 6-9. The best results for every function are emboldened. Table 10 presents the Friedman test results. The Wilcoxon's rank-sum test results are shown in Table 11.

Table 6 shows the optimal solutions for each uni-modal function, which can be used to analyze the algorithms' exploitation capabilities. In terms of standard deviation and average fitness, the SLSMA is ranked second in the category for $F_1$ and $F_3$. This shows that the the SLSMA has the necessary exploitation capabilities to perform well on these functions. Moreover, the novel search algorithm in the SLSMA allows it to find better solutions than the SMA.

Table 7 displays the results on the various functions of the simple multi-modal category, $F_4 - F_{10}$. The results of the experiments reveal that the SLSMA performs best on $F_5 - F_7$ and $F_{10}$. This is primarily due to the adoption of crossover operators and mutation. The DE performs better than the SLSMA on $F_4$ and $F_9$. Nevertheless, the latter still has an overall better exploration capabilities.

The hybrid functions, $F_{11} - F_{20}$, combine simple multi-modal and uni-modal functions. They are more challenging than the previous two groups. They have a better suitability for algorithmic exploration-exploitation balance testing. Table 8 shows that the SLSMA performs better than the other algorithms on $F_{17}$, recording overall best fitness on $F_{11} - F_{13}$, and $F_{15}$. The SLSMA ranks third in the hybrid functions group, but it is able to balance exploration and exploitation better than the other algorithm. This is evidenced by its attainment of best fitness in most functions.

The composition functions, $F_{21} - F_{30}$, constitute a combination of the uni-modal, the simple multi-modal, and the hybrid functions in a nonlinear fashion. They are the most challenging functions in the test suite. They are also the most ideal for evaluating the overall algorithmic effectiveness. Table 9 shows the results on $F_{21} - F_{30}$ to evaluate the algorithmic effectiveness of the SLSMA. It shows that the SLSMA has the best performance on $F_{21}$, $F_{23}$, $F_{25}$, $F_{26}$, and $F_{29}$. The SLSMA again has the optimal average fitness on $F_{24}$, $F_{27}$, and $F_{30}$, demonstrating a solid evidence for the SLSMA's effectiveness in the search for optimal fitness in composition functions. This result also highlights the pivotal roles of the new search mechanism, the dynamic switch operator, and the dynamic perturbation technique in the SLSMA's capacity to escape local optima and premature convergence.

According to the results of the Friedman test, the SLSMA is the most likely to perform well in all the functions evaluated. It was followed by DE and SMA. Table 11 shows the comparison of the various averages of the algorithms' best solutions. The significance level for the test is set at 5%, which indicates that the results are statistically significant. The results of the test indicate that the SLSMA's performance is robust and does not depend on random chance. This is because the algorithms' optimal solutions were developed through a systematic process.

**5.1.2. Analysis of Algorithmic Convergence Behaviour**

The five best-performing algorithms from the analysis of the algorithmic effectiveness were further examined for their convergence performance on the $F_1 - F_{10}$ functions. The respective algorithmic convergence curves are shown in Figure 3. The remainder of the algorithmic convergence performance curves of the algorithm on $F_{11} - F_{30}$ are presented in Appendix I.

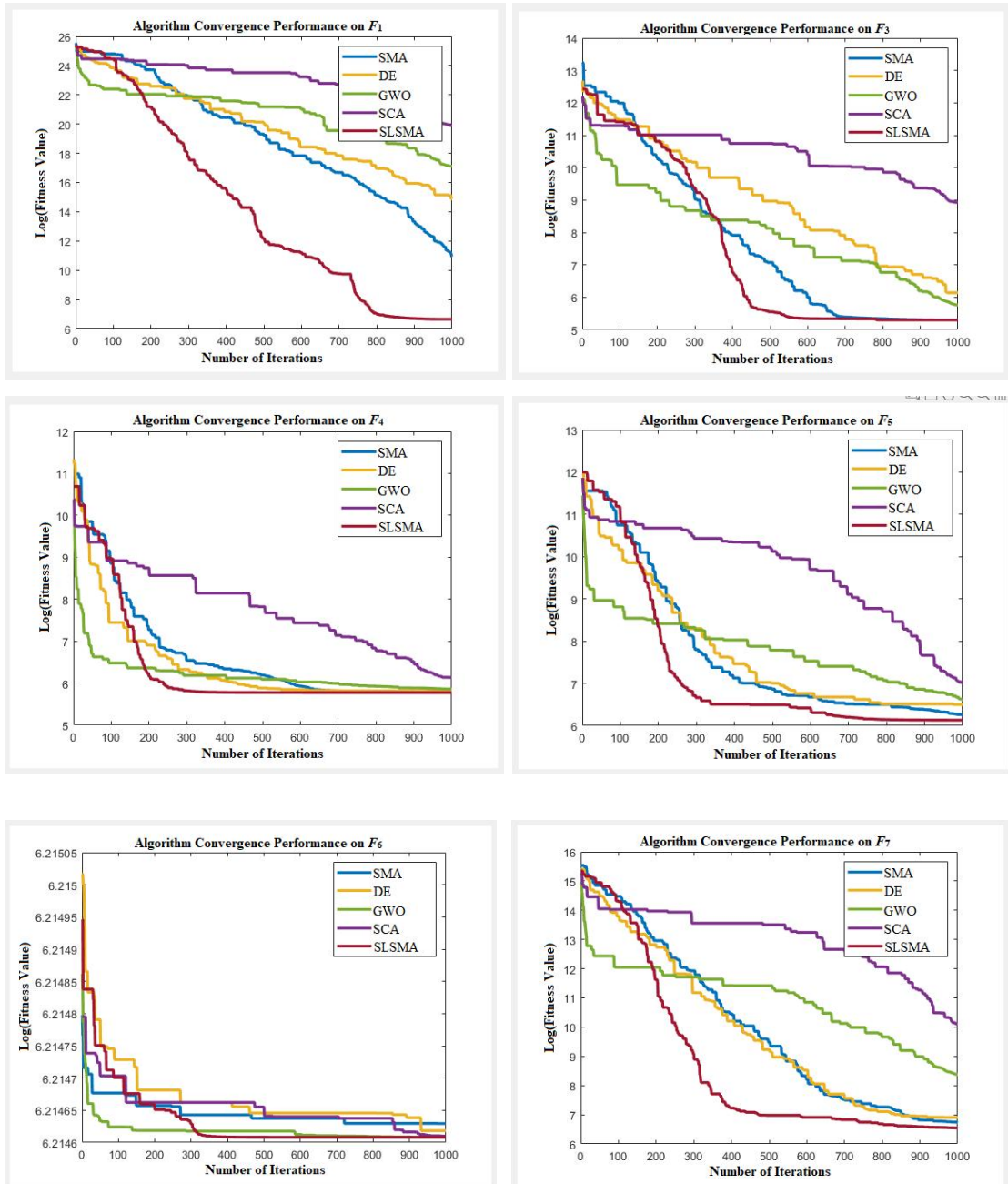

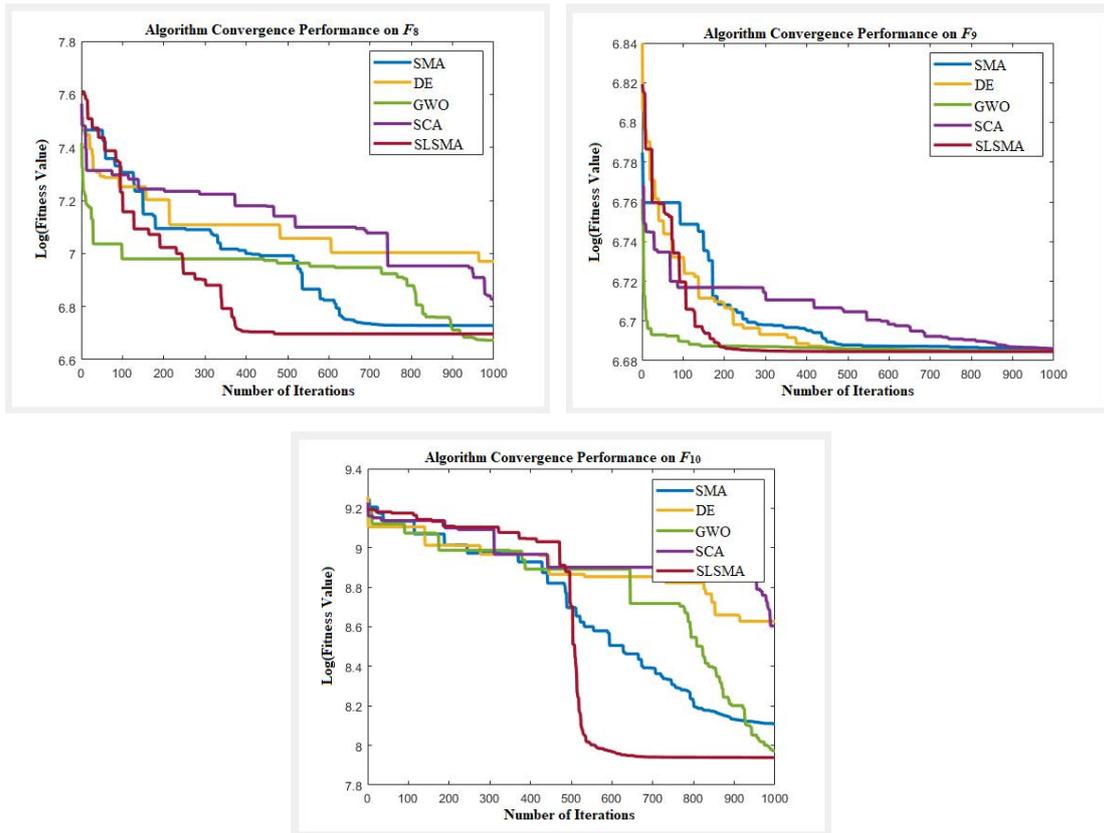

*Figure 3:* Convergence curves of five algorithms on the benchmark functions, $F_1$ to $F_{10}$

**Table 6: Results of Experiments on Congress on Evolutionary Computation Uni-modal Functions**

| Function | Metric | DE | GWO | PSO | SCA | SLSMA | SMA | SSA | WOA |
|---|---|---|---|---|---|---|---|---|---|
| $F_1$ | Avg | 8.3174E+06 | 3.9045E+08 | 3.1114E+10 | 4.5858E+08 | 5.2162E+04 | **4.9315E+04** | 3.6401E+08 | 5.3789E+09 |
| | Best | 2.9284E+06 | 2.7352E+07 | 4.7688E+10 | 4.4536E+08 | **7.8395E+02** | 5.7042E+04 | 3.3148E+06 | 3.1317E+09 |
| | Std | 1.1897E+07 | 7.6667E+08 | 3.1622E+10 | 5.1450E+08 | 1.2405E+05 | **5.5517E+04** | 2.5827E+09 | 6.4770E+09 |
| $F_3$ | Avg | 1.0575E+03 | 2.0234E+03 | 4.7688E+04 | 4.9722E+03 | 1.2812E+02 | **1.0981E+02** | 9.3952E+03 | 1.9523E+04 |
| | Best | 4.7180E+02 | 3.2334E+02 | 4.9518E+04 | 7.6667E+03 | **2.0336E+02** | 2.0336E+02 | 1.3625E+03 | 2.1861E+04 |
| | Std | 1.3523E+03 | 3.5791E+03 | 4.9111E+04 | 5.1450E+03 | 1.3320E+02 | **9.4969E+01** | 2.2370E+04 | 2.0336E+04 |

**Table 7: Results of Experiments on Congress on Evolutionary Computation Simple Multi-modal Functions**

| Function | Metric | DE | GWO | PSO | SCA | SLSMA | SMA | SSA | WOA |
|---|---|---|---|---|---|---|---|---|---|
| $F_4$ | Avg | **1.88E+02** | 2.63E+02 | 7.39E+03 | 2.84E+02 | 1.97E+02 | 1.91E+02 | 1.23E+03 | 6.29E+02 |
| | Best | 3.53E+02 | 3.72E+02 | 9.75E+03 | 4.89E+02 | **3.41E+02** | 3.53E+02 | 4.79E+02 | 7.17E+02 |
| | Std | **1.73E+02** | 2.64E+02 | 7.67E+03 | 2.69E+02 | 1.85E+02 | 1.76E+02 | 2.84E+03 | 6.91E+02 |
| $F_5$ | Avg | 3.76E+02 | 9.39E+02 | 3.92E+04 | 9.32E+02 | **2.78E+02** | 3.14E+02 | 6.12E+03 | 1.32E+04 |
| | Best | 7.06E+02 | 7.88E+02 | 4.97E+04 | 1.19E+03 | **4.87E+02** | 5.51E+02 | 6.91E+03 | 1.35E+04 |
| | Std | 3.61E+02 | 1.34E+03 | 4.01E+04 | 9.64E+02 | **2.64E+02** | 3.00E+02 | 6.56E+03 | 1.40E+04 |
| $F_6$ | Avg* | 2.73E+02 | 2.73E+02 | 2.73E+02 | 2.73E+02 | **2.73E+02** | 2.73E+02 | 2.73E+02 | 2.73E+02 |
| | Best* | 5.31E+02 | 5.31E+02 | 5.31E+02 | 5.31E+02 | **5.31E+02** | 5.31E+02 | 5.31E+02 | 5.31E+02 |
| | Std* | 2.58E+02 | 2.58E+02 | 2.58E+02 | 2.58E+02 | **2.58E+02** | 2.58E+02 | 2.58E+02 | 2.58E+02 |
| $F_7$ | Avg | 7.09E+02 | 2.03E+04 | 1.30E+06 | 1.93E+04 | **4.17E+02** | 4.93E+02 | 3.34E+04 | 2.14E+05 |
| | Best | 1.06E+03 | 4.62E+03 | 1.69E+06 | 2.58E+04 | **7.39E+02** | 8.88E+02 | 1.40E+04 | 1.42E+05 |
| | Std | 7.35E+02 | 4.48E+04 | 1.32E+06 | 2.04E+04 | **4.03E+02** | 4.79E+02 | 1.56E+05 | 2.42E+05 |
| $F_8$ | Avg | 6.22E+02 | 4.95E+02 | 8.02E+02 | 5.37E+02 | **4.82E+02** | 5.35E+02 | 8.22E+02 | 8.02E+02 |
| | Best | 1.13E+03 | **8.39E+02** | 1.43E+03 | 9.80E+02 | 8.60E+02 | 8.88E+02 | 1.43E+03 | 1.25E+03 |
| | Std | 6.07E+02 | 4.88E+02 | 7.88E+02 | 5.23E+02 | **4.68E+02** | 5.23E+02 | 8.08E+02 | 7.92E+02 |
| $F_9$ | Avg* | **4.33E+02** | 4.33E+02 | 4.57E+02 | 4.33E+02 | 4.34E+02 | 4.37E+02 | 4.39E+02 | 4.44E+02 |
| | Best* | 8.49E+02 | 8.49E+02 | 8.79E+02 | 8.51E+02 | **8.49E+02** | 8.51E+02 | 8.59E+02 | 8.60E+02 |

|  | Std | **4.17E+02** | 4.18E+02 | 4.41E+02 | 4.18E+02 | 4.19E+02 | 4.22E+02 | 4.24E+02 | 4.29E+02 |
|  | Avg | 4.17E+03 | 2.82E+03 | 4.52E+03 | 3.28E+03 | **2.18E+03** | 2.34E+03 | 4.22E+03 | 3.53E+03 |
| $F_{10}$ | Best | 5.94E+03 | 3.08E+03 | 8.31E+03 | 5.79E+03 | **2.97E+03** | 3.54E+03 | 4.47E+03 | 4.84E+03 |
|  | Std | 4.20E+03 | 3.20E+03 | 4.51E+03 | 3.28E+03 | **2.22E+03** | 2.37E+03 | 4.31E+03 | 3.63E+03 |

Table 8: Results of Experiments on Congress on Evolutionary Computation Hybrid Functions

| Function | Metric | DE | GWO | PSO | SCA | SLSMA | SMA | SSA | WOA |
|---|---|---|---|---|---|---|---|---|---|
|  | Avg | 2.97E+04 | 5.97E+04 | 7.83E+08 | 5.97E+04 | 5.36E+04 | **2.63E+04** | 5.80E+07 | 4.13E+06 |
| $F_{11}$ | Best | 2.40E+04 | 6.54E+04 | 4.56E+08 | 5.97E+04 | **1.37E+04** | 1.76E+04 | 5.52E+04 | 6.75E+04 |
|  | Std | 3.96E+04 | 6.53E+04 | 9.28E+08 | 6.39E+04 | 7.35E+04 | **3.00E+04** | 2.75E+08 | 3.12E+07 |
|  | Avg | 4.98E+06 | 1.25E+07 | 4.02E+09 | 1.90E+07 | 2.26E+06 | **3.74E+05** | 7.77E+08 | 2.05E+07 |
| $F_{12}$ | Best | 3.51E+06 | 4.06E+06 | 4.73E+09 | 2.10E+07 | 2.10E+05 | **5.91E+04** | 6.16E+05 | 1.38E+06 |
|  | Std | 7.64E+06 | 1.73E+07 | 4.17E+09 | 2.01E+07 | 3.94E+06 | **5.43E+05** | 1.83E+09 | 3.41E+07 |
|  | Avg | 1.15E+05 | 4.24E+06 | 4.49E+09 | 6.26E+06 | 1.85E+05 | **6.77E+04** | 3.30E+08 | 4.24E+07 |
| $F_{13}$ | Best | 3.04E+04 | 3.51E+05 | 4.05E+09 | 6.72E+06 | **1.15E+04** | 3.59E+04 | 5.08E+04 | 7.29E+05 |
|  | Std | 1.90E+05 | 1.09E+07 | 4.73E+09 | 7.41E+06 | 5.66E+05 | **9.47E+04** | 1.10E+09 | 1.03E+08 |
|  | Avg | **4.49E+03** | 2.13E+05 | 2.23E+06 | 1.77E+05 | 1.69E+05 | 1.38E+05 | 1.88E+06 | 8.11E+05 |
| $F_{14}$ | Best | **4.25E+03** | 8.11E+04 | 9.14E+05 | 1.09E+05 | 3.54E+04 | 7.43E+04 | 1.66E+05 | 1.74E+05 |
|  | Std | **5.54E+03** | 3.98E+05 | 2.87E+06 | 2.36E+05 | 2.41E+05 | 1.95E+05 | 2.91E+06 | 1.39E+06 |
|  | Avg | 7.97E+04 | 2.76E+05 | 2.07E+09 | 1.08E+06 | 6.42E+04 | **6.35E+04** | 7.12E+08 | 1.09E+07 |
| $F_{15}$ | Best | 4.89E+04 | 1.00E+05 | 1.14E+09 | 4.59E+05 | **1.23E+04** | 6.93E+04 | 3.89E+04 | 1.52E+05 |
|  | Std | 1.17E+05 | 5.60E+05 | 2.25E+09 | 1.35E+06 | 1.47E+05 | **7.88E+04** | 2.04E+09 | 3.61E+07 |
|  | Avg | **8.89E+02** | 7.58E+05 | 4.05E+08 | 2.30E+05 | 3.30E+03 | 2.50E+03 | 9.32E+04 | 1.32E+06 |
| $F_{16}$ | Best | 1.62E+03 | 2.08E+04 | 1.04E+08 | 2.33E+04 | 3.27E+03 | **1.61E+03** | 3.20E+03 | 1.25E+04 |
|  | Std | **8.79E+02** | 1.41E+06 | 6.18E+08 | 5.53E+05 | 4.36E+03 | 3.95E+03 | 2.11E+05 | 3.06E+06 |
|  | Avg | 3.07E+03 | 2.91E+04 | 8.93E+11 | 2.09E+04 | **1.65E+03** | 1.69E+03 | 1.35E+14 | 1.44E+08 |
| $F_{17}$ | Best | 2.62E+03 | 1.53E+04 | 2.01E+09 | 1.99E+04 | **2.21E+03** | 2.29E+03 | 2.16E+04 | 2.47E+04 |
|  | Std | 3.78E+03 | 3.30E+04 | 2.99E+12 | 2.20E+04 | **1.67E+03** | 1.71E+03 | 8.93E+14 | 1.10E+09 |
|  | Avg | **2.58E+04** | 1.99E+05 | 6.37E+06 | 1.66E+05 | 8.89E+04 | 6.14E+04 | 2.81E+06 | 1.26E+05 |
| $F_{18}$ | Best | **1.75E+04** | 5.74E+04 | 2.37E+06 | 1.25E+05 | 3.07E+04 | 6.36E+04 | 5.10E+04 | 5.45E+04 |
|  | Std | **3.26E+04** | 2.81E+05 | 8.45E+06 | 1.97E+05 | 1.18E+05 | 6.63E+04 | 7.20E+06 | 1.84E+05 |

| | Avg | **1.50E+03** | 4.06E+07 | 1.18E+12 | 7.20E+06 | 3.92E+05 | 2.65E+04 | 8.04E+08 | 4.32E+07 |
|---|---|---|---|---|---|---|---|---|---|
| $F_{19}$ | Best | **2.03E+03** | 3.14E+04 | 1.82E+10 | 1.19E+06 | 8.64E+03 | 1.05E+04 | 1.51E+04 | 7.39E+04 |
| | Std | **1.62E+03** | 1.24E+08 | 2.71E+12 | 1.22E+07 | 7.85E+05 | 3.50E+04 | 3.65E+09 | 1.32E+08 |
| | Avg | 1.35E+03 | 1.33E+03 | 5.94E+03 | 1.56E+03 | 1.36E+03 | **1.22E+03** | 8.65E+03 | 6.38E+03 |
| $F_{20}$ | Best | 2.14E+03 | 2.08E+03 | 6.78E+03 | 2.53E+03 | 2.13E+03 | **2.07E+03** | 6.99E+03 | 6.37E+03 |
| | Std | 1.37E+03 | 1.34E+03 | 6.05E+03 | 1.56E+03 | 1.36E+03 | **1.22E+03** | 9.53E+03 | 6.89E+03 |

Table 9: Results of Experiments on Congress on Evolutionary Computation Composition Functions

| Function | Metric | DE | GWO | PSO | SCA | SLSMA | SMA | SSA | WOA |
|---|---|---|---|---|---|---|---|---|---|
| | Avg* | 1.41E+03 | 1.74E+03 | 1.57E+04 | 1.78E+03 | **1.18E+03** | 1.18E+03 | 9.16E+03 | 8.73E+03 |
| $F_{21}$ | Best* | 2.30E+03 | 2.53E+03 | 1.60E+04 | 3.00E+03 | **2.24E+03** | 2.24E+03 | 2.54E+03 | 4.23E+03 |
| | Std* | 1.43E+03 | 1.84E+03 | 1.72E+04 | 1.79E+03 | **1.17E+03** | 1.17E+03 | 1.32E+04 | 1.09E+04 |
| | Avg* | 1.22E+03 | 1.22E+03 | 2.03E+03 | 1.25E+03 | 1.21E+03 | **1.21E+03** | 2.38E+03 | 2.09E+03 |
| $F_{22}$ | Best | 2.41E+03 | **2.38E+03** | 3.13E+03 | 2.45E+03 | 2.40E+03 | 2.40E+03 | 2.67E+03 | 3.02E+03 |
| | Std* | 1.21E+03 | 1.21E+03 | 2.03E+03 | 1.23E+03 | 1.20E+03 | **1.20E+03** | 2.56E+03 | 2.13E+03 |
| | Avg | 3.03E+03 | 2.67E+03 | 2.86E+04 | 3.22E+03 | **1.31E+03** | 1.33E+03 | 1.30E+04 | 1.15E+04 |
| $F_{23}$ | Best | 2.78E+03 | 2.93E+03 | 3.49E+04 | 3.96E+03 | **2.54E+03** | 2.59E+03 | 2.87E+03 | 9.45E+03 |
| | Std | 4.03E+03 | 3.12E+03 | 2.90E+04 | 3.28E+03 | **1.30E+03** | 1.32E+03 | 1.90E+04 | 1.35E+04 |
| | Avg | 1.81E+03 | 2.27E+03 | 1.78E+04 | 2.41E+03 | **1.35E+03** | 1.37E+03 | 4.71E+03 | 8.66E+03 |
| $F_{24}$ | Best | 2.78E+03 | 3.00E+03 | 2.60E+04 | 4.12E+03 | 2.65E+03 | **2.57E+03** | 2.82E+03 | 5.78E+03 |
| | Std | 2.12E+03 | 2.57E+03 | 1.82E+04 | 2.41E+03 | **1.34E+03** | 1.36E+03 | 7.81E+03 | 1.01E+04 |
| | Avg | 1.51E+03 | 1.55E+03 | 4.33E+03 | 1.58E+03 | **1.48E+03** | 1.51E+03 | 1.99E+03 | 1.84E+03 |
| $F_{25}$ | Best | 2.99E+03 | 3.02E+03 | 6.69E+03 | 3.08E+03 | **2.93E+03** | 2.99E+03 | 3.28E+03 | 3.27E+03 |
| | Std | 1.50E+03 | 1.53E+03 | 4.39E+03 | 1.57E+03 | **1.47E+03** | 1.50E+03 | 2.04E+03 | 1.83E+03 |
| | Avg | 1.78E+03 | 1.82E+03 | 4.26E+03 | 1.88E+03 | **1.56E+03** | 1.79E+03 | 6.50E+03 | 2.53E+03 |
| $F_{26}$ | Best | 3.54E+03 | 3.55E+03 | 6.00E+03 | 3.61E+03 | **3.09E+03** | 3.54E+03 | 4.59E+03 | 3.61E+03 |
| | Std | 1.77E+03 | 1.79E+03 | 4.37E+03 | 1.86E+03 | **1.54E+03** | 1.77E+03 | 7.89E+03 | 2.85E+03 |
| | Avg | 1.67E+03 | 1.68E+03 | 2.11E+03 | 1.74E+03 | **1.66E+03** | 1.67E+03 | 2.29E+03 | 2.06E+03 |
| $F_{27}$ | Best* | **3.29E+03** | 3.30E+03 | 3.72E+03 | 3.42E+03 | 3.29E+03 | 3.30E+03 | 3.96E+03 | 3.61E+03 |
| | Std | 1.65E+03 | 1.66E+03 | 2.10E+03 | 1.73E+03 | **1.64E+03** | 1.66E+03 | 2.29E+03 | 2.05E+03 |
| | Avg | **1.57E+03** | 1.69E+03 | 2.99E+03 | 1.70E+03 | 1.71E+03 | 1.58E+03 | 3.06E+03 | 1.85E+03 |
| $F_{28}$ | Best | 2.96E+03 | 3.19E+03 | 4.86E+03 | 3.22E+03 | 3.39E+03 | **2.89E+03** | 3.61E+03 | 3.42E+03 |
| | Std | **1.56E+03** | 1.67E+03 | 3.00E+03 | 1.69E+03 | 1.69E+03 | 1.57E+03 | 3.44E+03 | 1.86E+03 |
| | Avg | 5.57E+05 | 5.46E+06 | 5.37E+10 | 2.96E+07 | **2.34E+03** | 4.54E+04 | 3.64E+12 | 1.86E+08 |
| $F_{29}$ | Best | 1.71E+04 | 4.69E+04 | 3.90E+09 | 6.39E+05 | **3.73E+03** | 8.35E+03 | 2.43E+08 | 2.86E+05 |
| | Std | 2.85E+06 | 1.31E+07 | 1.08E+11 | 6.11E+07 | **2.34E+03** | 8.76E+04 | 2.45E+13 | 4.53E+08 |
| | Avg | 5.13E+05 | 1.90E+07 | 2.93E+10 | 2.00E+07 | **6.72E+04** | 8.51E+04 | 1.06E+11 | 9.90E+07 |
| $F_{30}$ | Best | 1.65E+05 | 2.01E+05 | 9.57E+09 | 1.79E+06 | **3.39E+03** | 3.60E+04 | 3.90E+05 | 5.09E+05 |
| | Std | 1.35E+06 | 1.13E+08 | 6.40E+10 | 6.73E+07 | 1.95E+05 | **1.24E+05** | 7.70E+11 | 2.60E+08 |

* The actual values vary

Table 10: Results of the Friedman Test

| Test | DE | GWO | PSO | SCA | SLSMA | SMA | SSA | WOA |
|---|---|---|---|---|---|---|---|---|
| Friedman Mean Rank | 1.900622 | 3.1854 | 7.103442 | 4.077312 | 0.976856 | 1.359104 | 5.616922 | 5.478888 |
| Rank | 3 | 4 | 8 | 5 | 1 | 2 | 7 | 6 |

Table 11: Results of Wilcoxon's Rank-sum Test

| S/N | SLSMA | Test Statistic | P-value | Significance |
|---|---|---|---|---|
| 1 | DE | 110.4272 | 0.01369722 | True |
| 2 | GWO | 25.4832 | 0.00004863 | True |
| 3 | PSO | 0.0000 | 0.00000000 | True |
| 4 | SCA | 8.4944 | 0.00000009 | True |
| 5 | SMA | 102.9946 | 0.00850502 | True |
| 6 | SSA | 2.1236 | 0.00000501 | True |
| 7 | WOA | 0.0000 | 0.00000003 | True |

Figure 3 and Appendix I show that the SLSMA has a better algorithmic convergence stability for all the benchmark functions. Also, the SLSMA, compared with the other algorithms, has an early-stage faster algorithmic convergence on most of the benchmark functions. This establishes the SLSMA's ability, driven by the dynamic perturbation technique, to overcome algorithmic stagnation and to speed up algorithmic convergence. For the benchmark functions, $F_1$, $F_7$, $F_{10}$, $F_{13}$, $F_{15}$, $F_{29}$, and $F_{30}$, the SLSMA achieves the minimum fitness statistic and fastest algorithmic convergence speed, which are critical in situations demanding efficient real-time computations.

**5.2. Multi-Drone Flight Path Planning**

The planning problems for multi-drone paths are more complex than those for benchmark functions due to their non-linear cost and constraints. The ESMA [60], CO-SMA [61], and SMA-AGDE [62] have been chosen as the SLSMA's competitors, since they have been shown to perform excellently relative to popular meta-heuristic algorithms. The ESMA incorporates the Sine Cosine Algorithm to improve the performance of the SMA. The CO-SMA applies a chaotic opposition-based approach that blends the chaotic search method with a crossover-opposition strategy. By applying an adaptive guided differential evolution technique, the SMA-AGDE enhances the parametric capabilities of the SMA. Table 12 details the parameter settings for CO-SMA, ESMA, SLSMA, SMA-AGDE, and SMA as indicated in their respective original papers.

**Table 12. Parameter Settings of Algorithms for Multi-Drone Path Optimization**

| S/N | Algorithm | Parameter Settings |
|---|---|---|
| 1 | CO-SMA | $z = 0.03$, and $\beta_0 \in [0, 1]$ |
| 2 | ESMA | $z = 0.03$, $r_2 \in [0, 2\pi]$, and $r_3 \in [0, 2]$ |
| 3 | SLSMA | $Cr = 0.5$, and $\zeta = 0.1$ |
| 4 | SMA-AGDE | $CR_1 \in [0.05, 0.15]$, and $CR_2 \in [0.9, 1.0]$ |
| 5 | SMA | $z = 0.03$ |

In this paper, the dimension of the test case is $1000^3$, with 5 drones ($M = 5$) and 10 waypoints ($N = 10$). The drone flight starting and terminal points are (0, 0, 0) and (1000, 1000, 0) respectively. The cost function together with the constraints is defined in Section 2. Table 13 details the settings for the various cost components.

**Table 13: Parameter Settings for Cost Components**

| S/N | Parameter | Value |
|---|---|---|
| 1 | Maximum yaw angle, $\alpha_{max}$ | 45° |
| 2 | Maximum pitch angle, $\beta_{max}$ | 45° |
| 3 | Lower height limit, $z_{lb}$ | 10 |
| 4 | Upper height limit, $z_{ub}$ | 950 |
| 5 | Minimum safe distance between drones, $d_{min}$ | 5 |
| 6 | Cost coefficient, $q_{height}$ | 100 |
| 7 | Cost coefficient, $q_{yaw}$ | 100 |
| 8 | Cost coefficient, $q_{pitch}$ | 100 |
| 9 | Cost coefficient, $q_{collis\_obs}$ | 100 |
| 10 | Cost coefficient, $q_{collis\_drone}$ | 100 |
| 11 | Weight coefficient, $\omega_1$ | 100 |
| 12 | Weight coefficient, $\omega_2$ | 1 |
| 13 | Weight coefficient, $\omega_3$ | 2 |
| 14 | Weight coefficient, $\omega_4$ | 3 |
| 15 | Weight coefficient, $\omega_5$ | 3 |

Table 14 presents the settings for the three scenarios considered in this paper, with $N = 30$, $t_{max} = 500$, and 30 independent runs.

**Table 14: Coordinates Information for Multi-Drone Path Planning Problems**

| S/N | Center Coordinate ($x_c$, $y_c$) | Slope ($x_{sl}$, $y_{sl}$) | Height ($h_m$) |
|---|---|---|---|
| | **Scenario I** | | |
| 1 | (250, 200) | (110, 105) | 505 |
| 2 | (420, 800) | (90, 140) | 745 |
| 3 | (720, 340) | (150, 140) | 605 |
| | **Scenario II** | | |
| 1 | (200, 330) | (100, 110) | 405 |
| 2 | (430, 700) | (90, 140) | 785 |
| 3 | (200, 730) | (110, 120) | 390 |
| 4 | (710, 800) | (90, 140) | 505 |
| 5 | (790, 200) | (90, 80) | 325 |
| 6 | (570, 210) | (100, 110) | 545 |
| | **Scenario III** | | |
| 1 | (150, 660) | (68, 65) | 385 |
| 2 | (460, 270) | (110, 95) | 535 |
| 3 | (740, 80) | (80, 50) | 425 |
| 4 | (190, 740) | (70, 70) | 265 |
| 5 | (390, 690) | (80, 80) | 495 |
| 6 | (710, 710) | (90, 160) | 705 |
| 7 | (940, 340) | (45, 76) | 395 |
| 8 | (390, 890) | (50, 50) | 295 |
| 9 | (840, 890) | (60, 70) | 495 |
| 10 | (290, 110), | (50, 60) | 355 |

The results of the five algorithms are compared in Table 15. The average algorithmic convergence curves shown in Figure 4. Also, Figures 5-8 depict the optimal path-planning results for the CO-SMA and the SLSMA, in both 2D and 3D. The corresponding average algorithmic convergence curves for the SMA, ESMA, and SMA-AGDE are given in Appendix II. The drone flight starting and terminal points are marked by a red square and a blue star respectively. The continuity of the trajectory arcs in the two-dimensional plot is an important factor used to identify potential drone-collisions. The collision-free trajectories are apparent between obstacles and drones along the continuous curves. Conversely, any discontinuities point to possible points of collision.

The first scenario where three obstacles were encountered provided the SLSMA with enhanced robustness in its algorithmic performance. As shown in Table 15, this is evidenced by its second-highest fitness value and optimal standard deviation. This is corroborated by Figure 4(a) wherein, out of the five algorithms, the SLSMA ranked second. The trajectory results shown in Figure 8(a), Figure IIa(i), and Figure IIb(i) indicate that CO-SMA, ESMA, and SMA are associated with identical drone flight trajectories which are characterized by potential drone collisions, attributable to small safety distances. There are two breaks in the curves, indicating that all drones will eventually collide with some obstacles. In Figure IIa(iii) and Figure IIb(iii), the SMA-AGDE prevent drone-drone collisions, however it is challenged by two obstacle collisions. On the contrary, Figure 5(a) and Figure 6(a) reveal that, apart from the small collision involving drone 3, the SLSMA achieves no-collision trajectories for all the drones. So, the SLSMA is the best algorithm in Scenario I.

**Table 15: Results for Multi-Drone Flight Path Planning**

| Scenarios | Metrics | CO-SMA | ESMA | SLSMA | SMA-AGDE | SMA |
|---|---|---|---|---|---|---|
| I | AVG | **4.26E+03** | 5.78E+03 | 5.28E+03 | 5.65E+03 | 8.41E+03 |
|  | BEST | 2.45E+03 | **2.43E+03** | 4.18E+03 | 4.37E+03 | 3.28E+03 |
|  | STD | 1.06E+03 | 1.44E+03 | **5.04E+02** | 6.97E+02 | 1.95E+03 |
| II | AVG | 6.23E+03 | 8.27E+03 | **5.14E+03** | 5.95E+03 | 9.24E+03 |
|  | BEST | **2.09E+03** | 3.70E+03 | 4.08E+03 | 4.88E+03 | 2.64E+03 |
|  | STD | 2.07E+03 | 2.30E+03 | **5.10E+02** | 6.31E+02 | 2.16E+03 |
| III | AVG | **4.00E+03** | 6.01E+03 | 5.02E+03 | 5.79E+03 | 8.27E+03 |
|  | BEST | **2.41E+03** | 2.94E+03 | 3.31E+03 | 4.35E+03 | 3.80E+03 |
|  | STD | 1.11E+03 | 1.64E+03 | **5.78E+02** | 7.29E+02 | 1.69E+03 |

The second scenario presents more challenges and obstacles than the first scenario. Table 15 and Figure 4(b) show that the SLSMA has the best of performance in the second scenario. It has the best average fitness, standard deviation, and highest algorithmic convergence effectiveness. Moreover, Figure 5(b) and Figure 6(b) corroborate that the SLSMA has no-collision flight trajectories for all the drones. However, Figure 8b, Figure IIc(i-iii) and Figure IId(i-iii) indicate that the drone flight trajectories from the competing algorithms are infeasible and characterized by potential collisions. So, the SLSMA is the best algorithm in the second scenario.

The third scenario is the most complex. It has the highest number of obstacles. It is also characterized by the obstacles with highest densities. Table 15 and Figure 4(c) reveal that, the SLSMA ranks second in average fitness, algorithmic convergence behaviour, and has the best standard deviation. This suggests that, in highly complex situations, the SLSMA still has a robust performance. Figure 7(c), Figure 8(c), Figure IIe(i-iii) and Figure IIf(i-iii) show that the CO-SMA, ESMA, SMA, and SMA-AGDE produce invalid drone flight trajectories, wherein the number of collisions increases as the scenario becomes more and more complex. Yet, Figure 5(c), and Figure 6(c) reveal that only the SLSMA generates complete collision-free drone flight trajectories. So, the SLSMA, again, is the best algorithm in the third scenario, Scenario III.

Overall, the results of the experiments show that the SLSMA has the best, stable, and consistent performance in terms of algorithmic effectiveness and convergence for the scenarios considered in this study.

The experimental results also reveal that, compared with the other competing algorithms, the SLSMA produces, overall, the most optimal drone flight trajectories. These findings cement the robust algorithmic performance, and suitability of the SLSMA for problems involving multi-drone path planning.

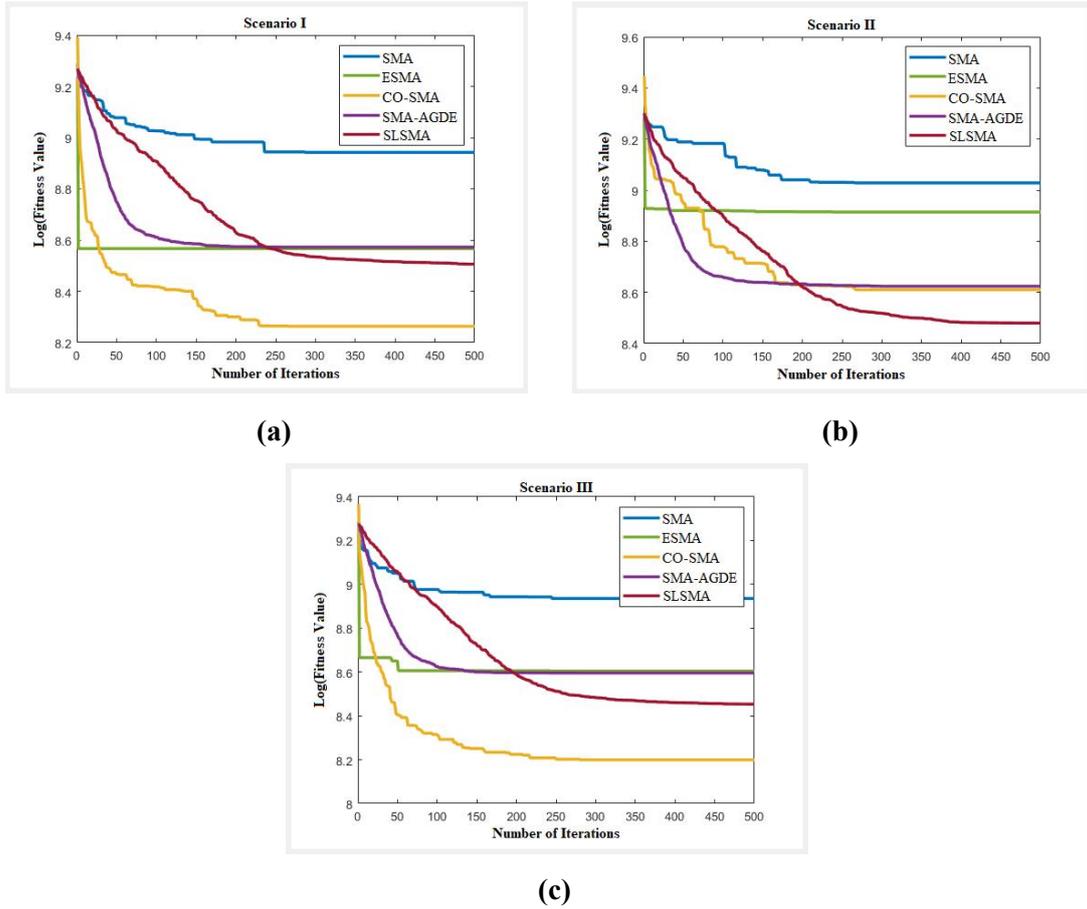

*Figure 4*: Convergence curves of selected algorithms for multi-drone path planning in **(a)** Scenario I **(b)** Scenario II, and **(c)** Scenario III.

Put differently, the experimental findings further indicate that, in comparison with alternative algorithms, the SLSMA generates the most optimal drone flight trajectories in the aggregate. These results substantiate the robust performance of the algorithm and confirm its suitability for application in multi-drone path planning problems.

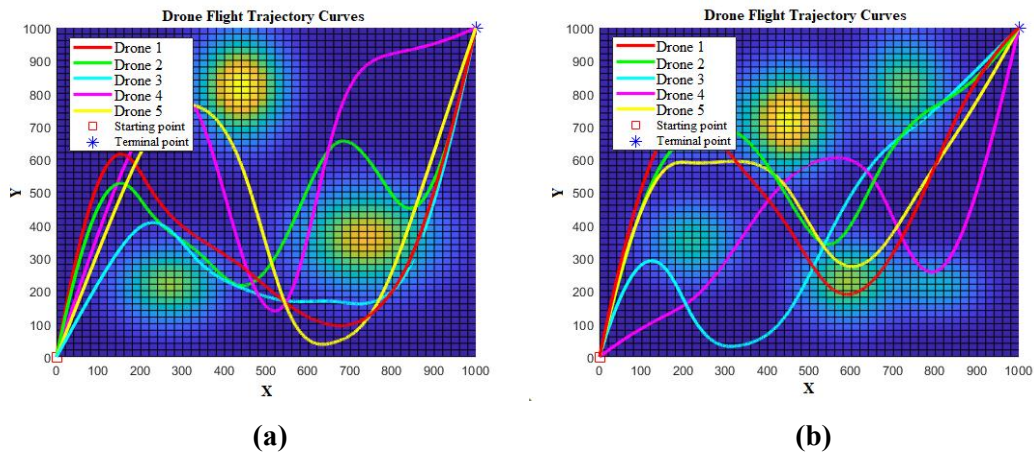

*Figure 5*: Two-dimensional view of drone flight trajectories using SLSMA in problem scenarios **(a)** Scenario I, **(b)** Scenario II, and **(c)** Scenario III.

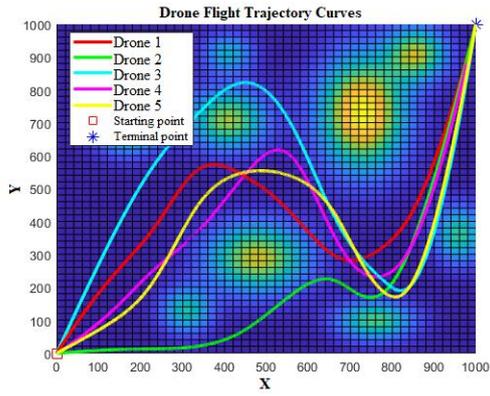
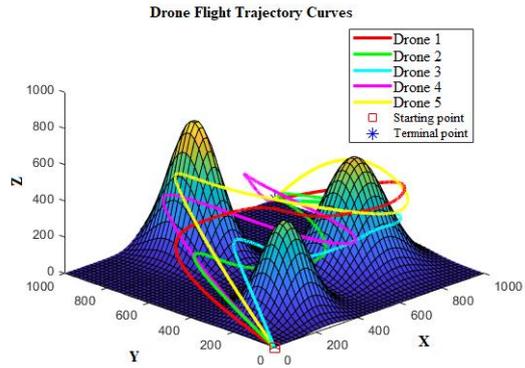

**(c)**                                             **(a)**

*Figure 5*: Cont.                         *Figure 6*: Cont.

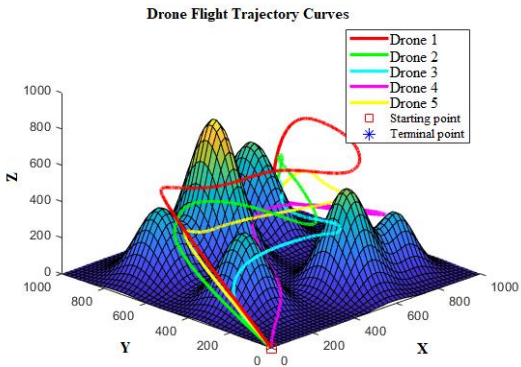
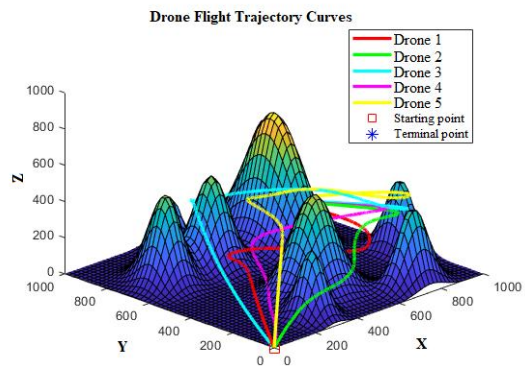

**(b)**                                             **(c)**

*Figure 6*: 3D view of drone flight trajectories using SLSMA in problem scenarios **(a)** Scenario I, **(b)** Scenario II, and **(c)** Scenario III.

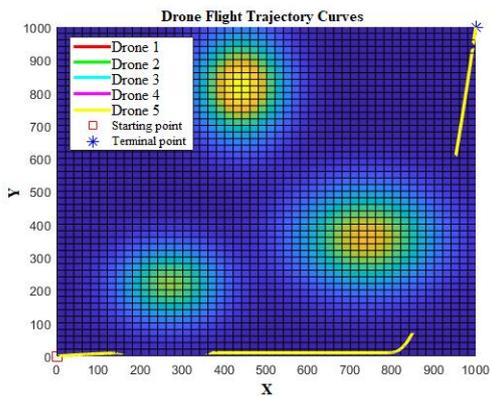
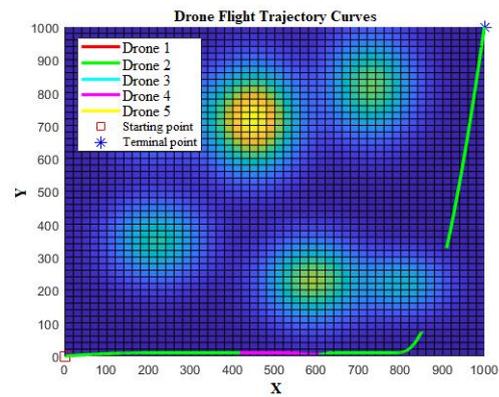

**(a)**                                             **(b)**

*Figure 7*: 2D view of drone flight trajectories using CO-SMA in problem scenarios **(a)** Scenario I, **(b)** Scenario II, and **(c)** Scenario III.

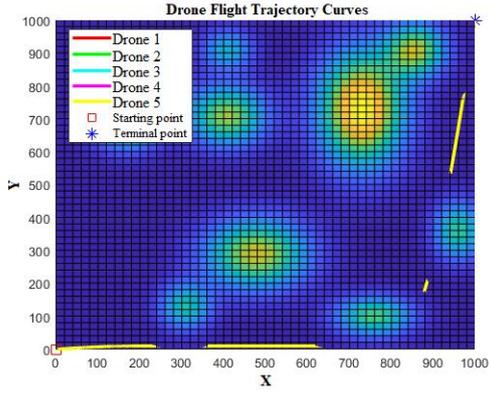
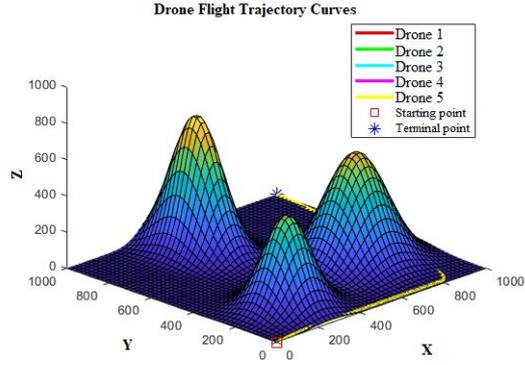

**(c)**                                       **(a)**

*Figure 7*: Cont..                 *Figure 8*: Cont..

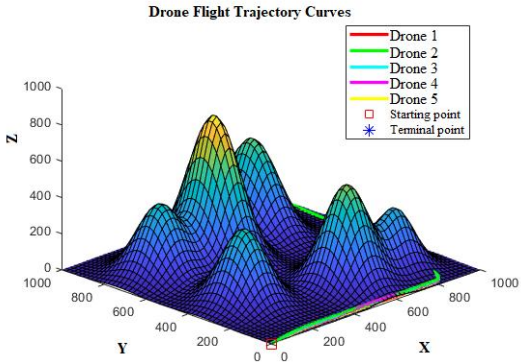
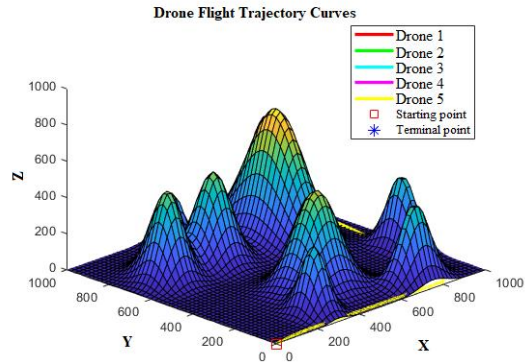

**(b)**                                       **(c)**

*Figure 8*: 3D view of drone flight trajectories using CO-SMA in problem scenarios **(a)** Scenario I, **(b)** Scenario II, and **(c)** Scenario III.

## 6. Conclusion

This paper has introduced a Self-Learning Slime Mould Algorithm (SLSMA), a novel meta-cognitive swarm intelligence framework designed to address the critical challenge of resilient trajectory optimization for UAV swarms in complex, cluttered environments. The proposed SLSMA fundamentally transcends the limitations of the traditional SMA by embedding self-directed learning and adaptive recovery mechanisms into its core architecture.

The algorithmic superiority of the SLSMA is anchored in three key innovations: a situation-aware search strategy that synergistically merges the global exploration prowess of a ranking-based differential evolution with the local exploitation focus of the SMA, achieving a dynamic and precise balance throughout the search process; a dynamic switching operator that replaces static parameters to autonomously maintain population diversity, thereby directly countering the issue of premature convergence; and an adaptive perturbation technique that acts as a meta-cognitive recovery mechanism, enabling the algorithm to intelligently identify and escape local optima, thus ensuring robust convergence to high-quality solutions. Furthermore, the application of cubic B-spline curves guarantees the generation of kinematically feasible and smooth flight trajectories suitable for real-world drone deployment.

Rigorous validation on the CEC 2017 benchmark suite and complex 3D path planning scenarios unequivocally demonstrates that the SLSMA outperforms a suite of state-of-the-art metaheuristics. The results confirm that our framework does not merely find paths; it generates resilient trajectories, exhibiting superior convergence characteristics, higher success rates, and greater reliability in the most demanding scenarios. This establishes the SLSMA as a robust and effective solver for the complex multi-UAV trajectory optimization problem.

This research, while comprehensive, opens several promising avenues for future research. The current study focused on a single-objective formulation within a static environment. Future work will investigate the extension of the SLSMA's meta-cognitive principles to:

1) Dynamic and uncertain environments, where real-time obstacle movement and unpredictable threats require continuous re-planning.
2) Multi-objective optimization problems, balancing competing goals such as mission time, energy consumption, and risk exposure.
3) Hardware-in-the-loop (HIL) simulations and physical flight tests to validate the algorithm's performance and resilience in real-time robotic systems.

By providing a foundational framework for adaptive and resilient swarm intelligence, this research paves the way for the next generation of autonomous UAVs capable of persistent and reliable operation in critically denied and complex terrains.

# APPENDICES

## Appendix I

Appendix I shows the convergence curves of DE, GWO, SCA, SLSMA, and SMA for 30 runs on the CEC 2017 benchmark functions from $F_{11}$ to $F_{30}$ with 30 dimensions.

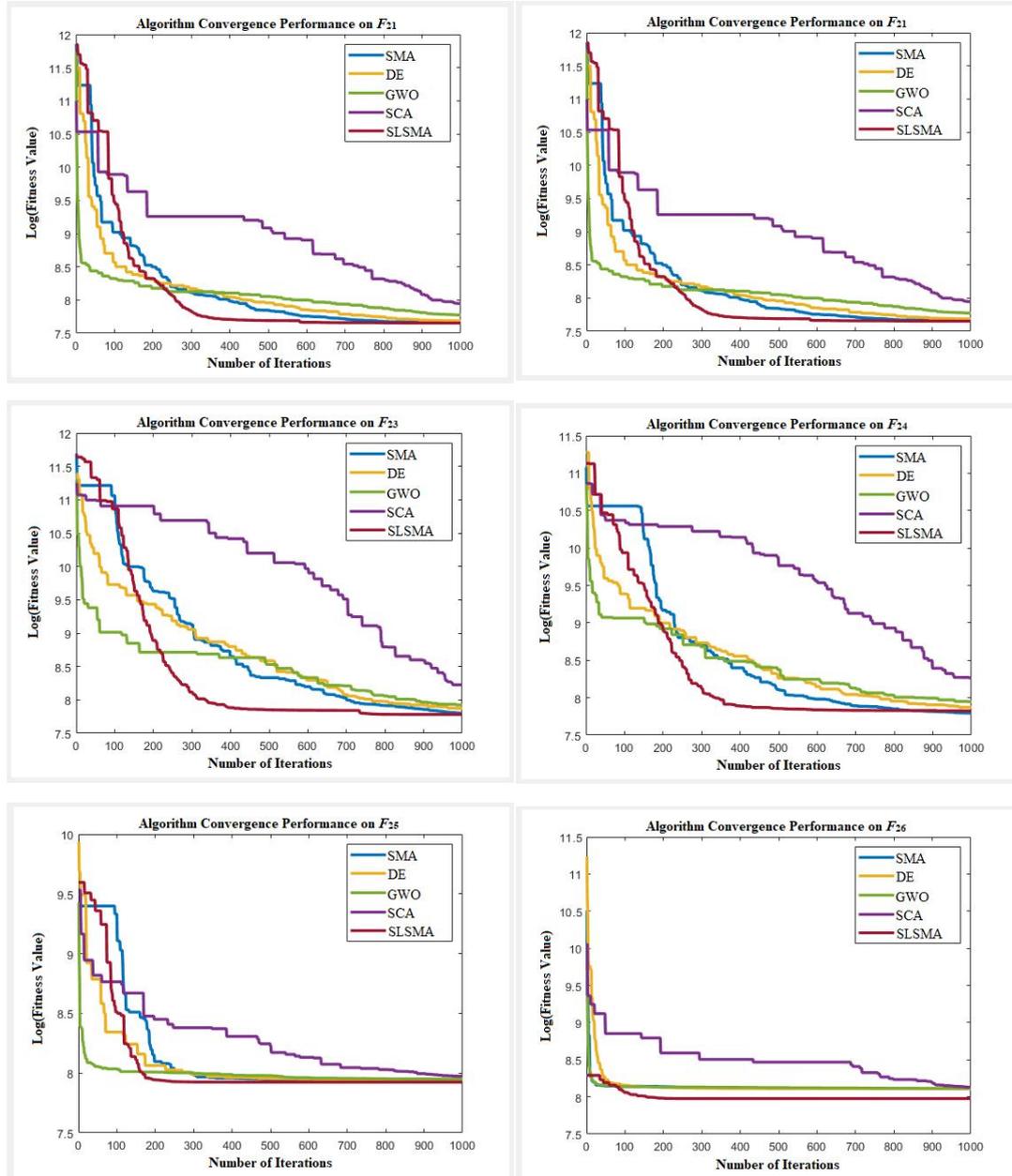

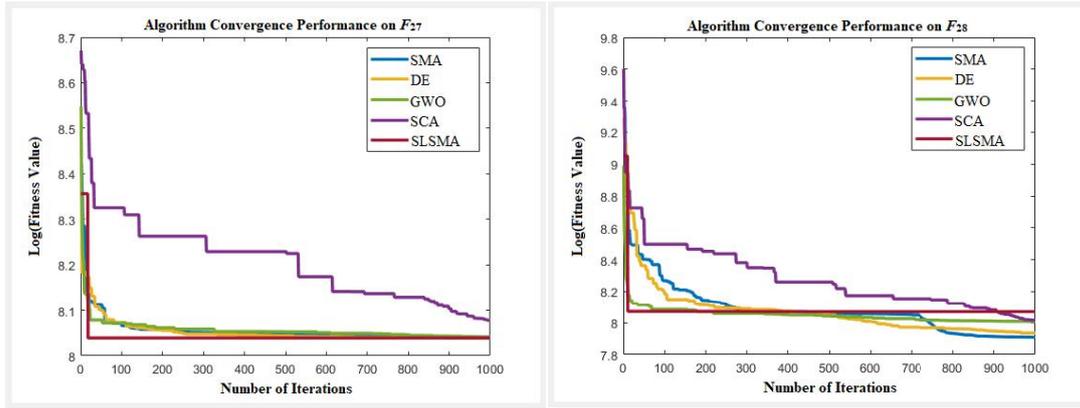

*Figure Ia*: Convergence curves of DE, GWO, SCA, SLSMA, and SMA on CEC 2017 functions

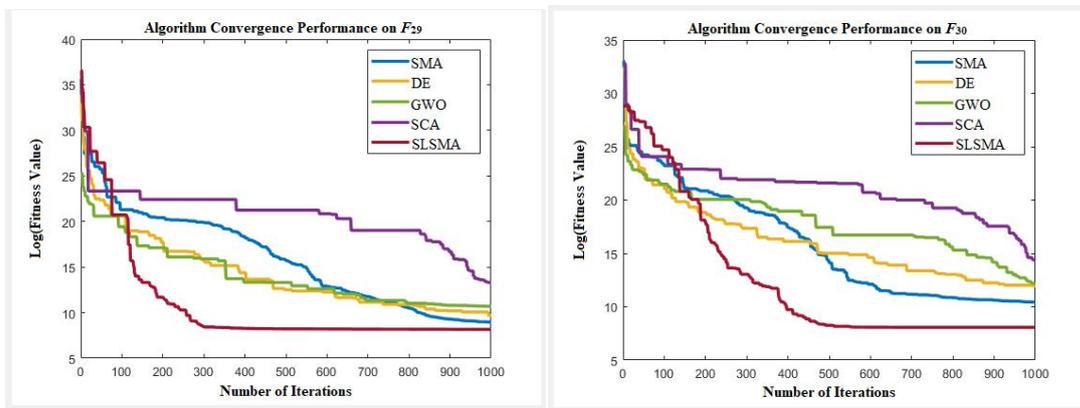

*Figure Ia*: Cont.

**Appendix II**

Appendix II presents the drone flight trajectories produced by SMA, ESMA, and SMA-AGDE for the three scenarios. Figure IIa, Figure IIb, and Figure IIc show the drone flight trajectories in Scenarios I, II, and III respectively.

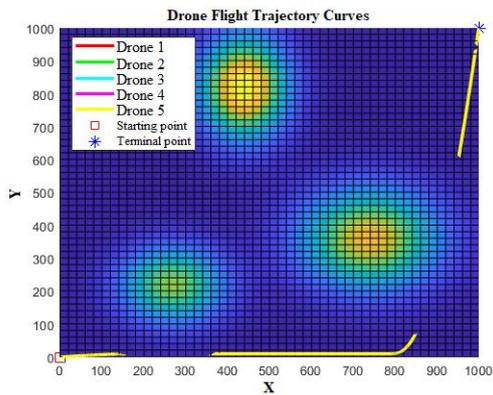 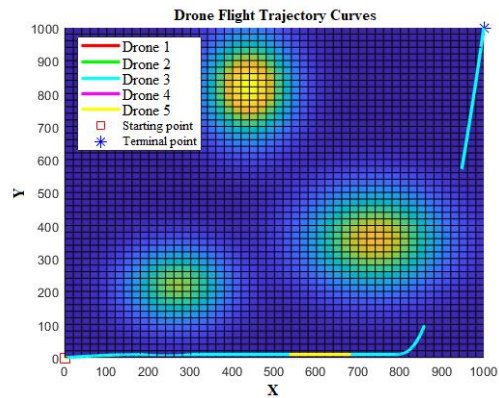

               **(i)**                             **(ii)**

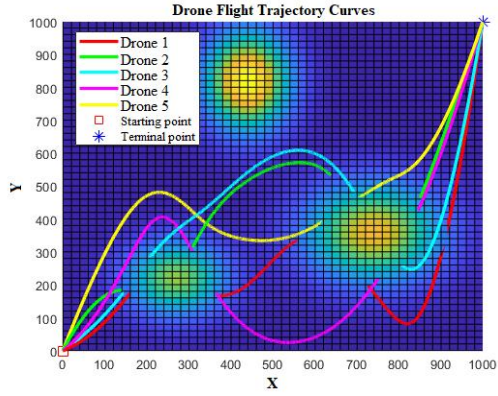

**(iii)**

*Figure IIa*: 2D view of drone flight trajectories using SMA, ESMA, and SMA-AGDE for problem scenario I. (a) SMA. (b) ESMA. (c) SMA-AGDE.

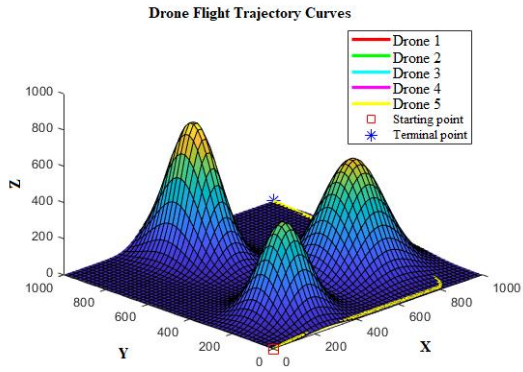 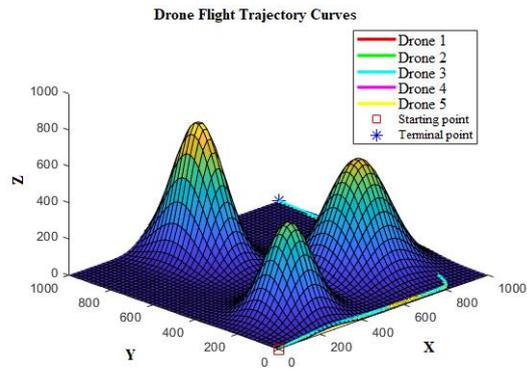

**(i)** **(ii)**

*Figure IIb*: 3D view of drone flight trajectories using SMA, ESMA, and SMA-AGDE for problem scenario I. (a) SMA. (b) ESMA. (c) SMA-AGDE

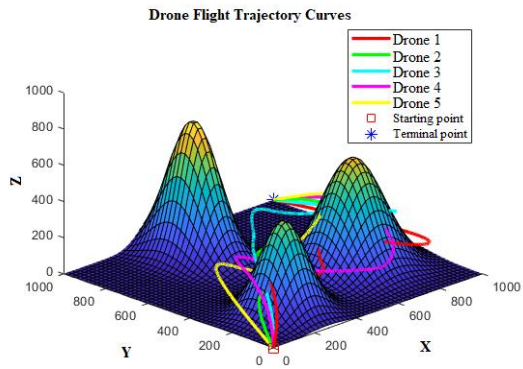 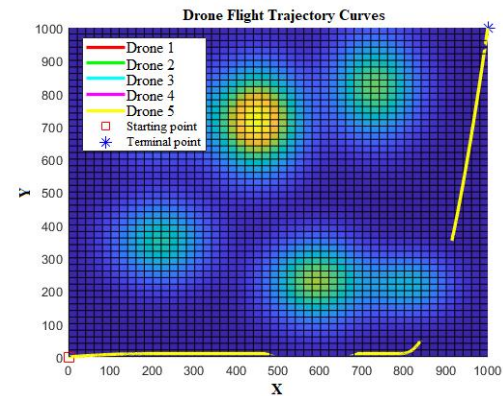

**(iii)** **(i)**

*Figure IIb*: Cont. *Figure IIc:* Cont.

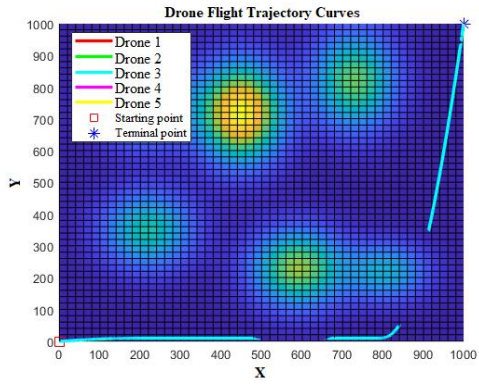 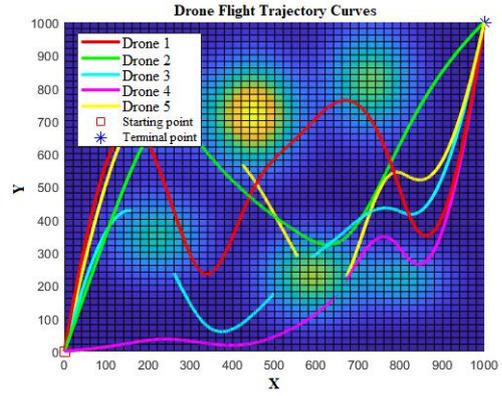

**(ii)** **(iii)**

*Figure IIc*: 2D view of drone flight trajectories using SMA, ESMA, and SMA-AGDE for problem scenario II. (a) SMA. (b) ESMA. **(c)** SMA-AGDE.

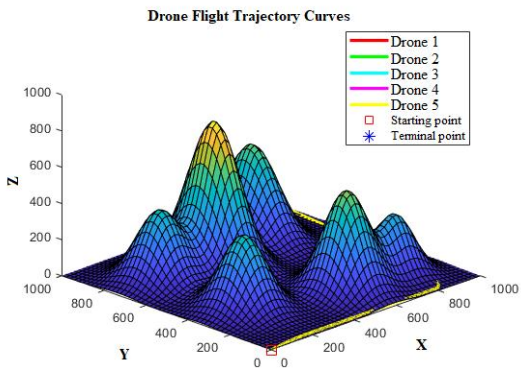 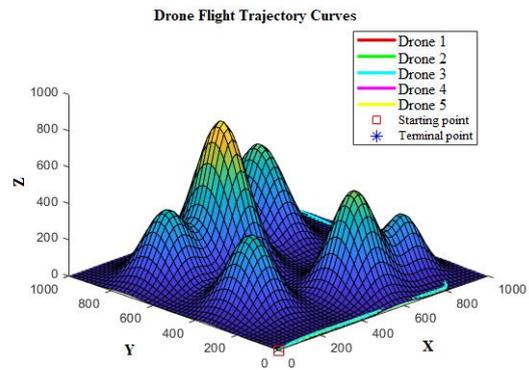

**(i)** **(ii)**

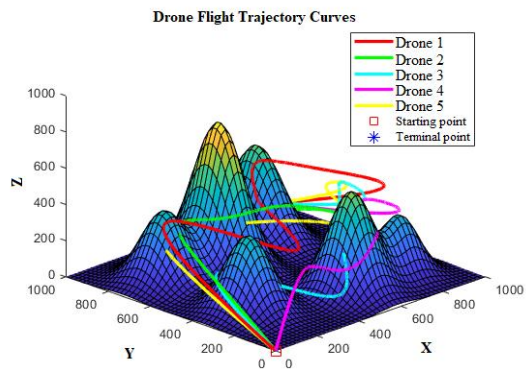

**(iii)**

*Figure IId*: 3D view of drone flight trajectories using SMA, ESMA, and SMA-AGDE for problem scenario II. (a) SMA. (b) ESMA. (c) SMA-AGDE.

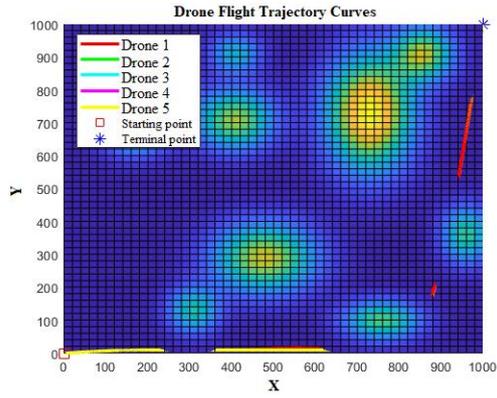 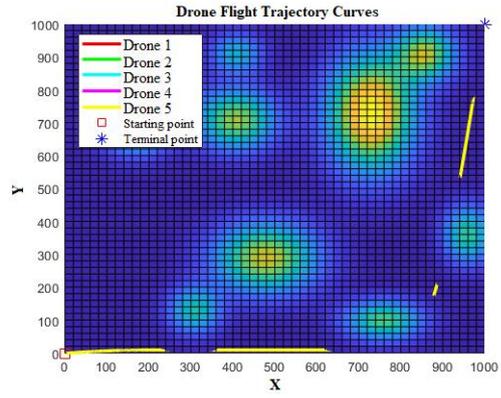

(i)                      (ii)

*Figure IIe*: 2D view of drone flight trajectories using SMA, ESMA, and SMA-AGDE for problem scenario III. (a) SMA. (b) ESMA. (c) SMA-AGDE.

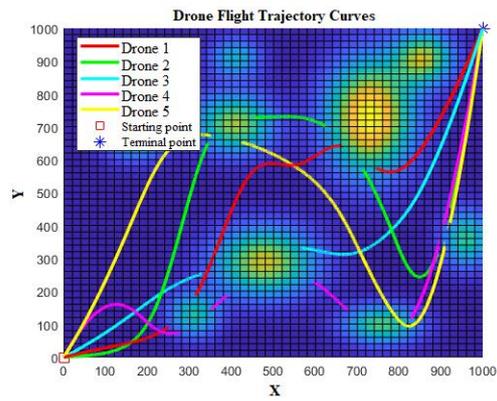 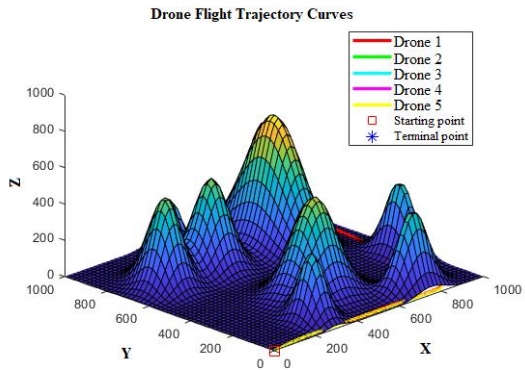

(iii)                   (i)

*Figure IIe*: Cont.               *Figure IIf*: Cont.

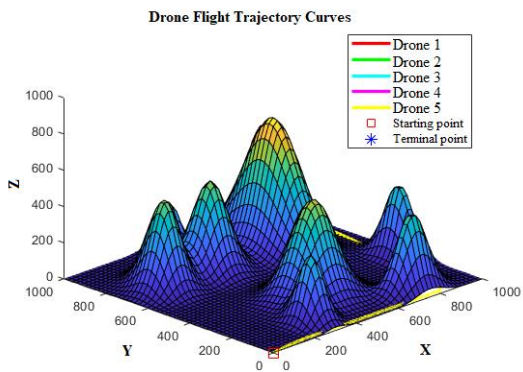 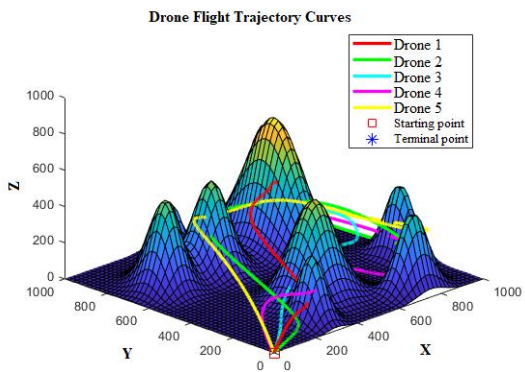

(ii)                    (ii)

*Figure IIf*: 3D view of drone flight trajectories using SMA, ESMA, and SMA-AGDE for problem scenario III. (a) SMA. (b) ESMA. (c) SMA-AGDE.